\documentclass[]{aa}

\usepackage{graphicx}
\usepackage{txfonts}
\usepackage[pdftex, colorlinks, linkcolor=red, citecolor=blue]{hyperref}
\usepackage{natbib}
\begin{document} 

\title{A model-based approach to the spatial and spectral calibration of NIRSpec onboard JWST}

\author{B. Dorner\inst{1}
\and G. Giardino\inst{2}
\and P. Ferruit\inst{2}
\and C. Alves de Oliveira\inst{3}
\and S. M. Birkmann\inst{3} 
\and T. B{\"o}ker\inst{3}
\and G. De Marchi\inst{2}
\and X. Gnata\inst{4}
\and J. K{\"o}hler\inst{4}
\and M. Sirianni\inst{3}
\and P. Jakobsen\inst{5}
}

\institute{Max-Planck-Institut f\"ur Astronomie, K\"onigstuhl 17, 69117 Heidelberg, Germany, \email{dorner@mpia.de}
\and European Space Agency, ESTEC, Postbus 299, 2200 AG Noordwijk, The Netherlands
\and European Space Agency, STScI, Baltimore, MD 21218, USA
\and Airbus Defence and Space, 81663 Munich, Germany
\and Dark Cosmology Centre, Niels Bohr Institute, University of Copenhagen, Juliane Maries Vej 30, 2100 Copenhagen \O, Denmark}

\date{Received 8 February 2016; accepted 26 April 2016}

\abstract
%[Context]
{The NIRSpec instrument for the James Webb Space Telescope
(JWST) can be operated in multiobject (MOS), long-slit, and integral
field (IFU) mode with spectral resolutions from 100 to 2700. Its MOS mode
uses about a quarter of a million individually addressable
minislits for object selection, covering a field of view of $\sim$9 $\mathrm{arcmin}^2$.}
%[Aims]
{The pipeline used to extract wavelength-calibrated spectra from NIRSpec detector images relies heavily on a model of NIRSpec optical geometry.
We demonstrate how dedicated calibration data from a small subset of NIRSpec modes and apertures can be used to optimize this parametric model to the necessary levels of fidelity.}
%[Methods] 
{Following an iterative procedure, the initial fiducial values of the model parameters
are manually adjusted and then automatically optimized, so that the
model predicted location of the images and spectral lines from the fixed slits,
the IFU, and a small subset of the MOS apertures matches their measured location in the main optical
planes of the instrument.}
%[Results] 
{The NIRSpec parametric model is able to reproduce the spatial and
  spectral position of the input spectra with high fidelity. The
  intrinsic accuracy (1-sigma, RMS) of the model, as measured from the
  extracted calibration spectra, is better than 1/10 of a
  pixel along the spatial direction and better than 1/20 of a
  resolution element in the spectral direction for all of the grating-based spectral modes.
  This is fully consistent with the corresponding allocation in the spatial and spectral calibration budgets of NIRSpec.}
%[Conclusion]
{}


\maketitle

%________________________________________________________________
\section{Introduction}

The James Webb Space Telescope (JWST) is a large near- and
mid-infrared space observatory with a primary mirror diameter of about
6.5~m, and is passively cooled to less than 50~K
\citep{Gardner:2006_00}. The spacecraft will be placed in an orbit
around the Sun-Earth Lagrange point L2 by an Ariane 5 rocket, whose
launch is scheduled for October 2018 from the space port of the
European Space Agency (ESA) in Kourou,
French Guyana. The JWST is a scientific project led by the National
Aeronautics and Space Administration (NASA), with major
contributions from ESA and the Canadian
Space Agency (CSA).  The observatory will carry a suite of four
science instruments, one of which is the Near Infrared Spectrograph
(NIRSpec), developed by ESA with
Airbus Defence and Space Germany (formerly EADS Astrium Germany GmbH)
as the prime contractor \citep{Bagnasco:2007_00}.

The primary goal of NIRSpec is to enable large spectroscopic surveys with JWST in the near-infrared with an emphasis on the study of the birth and assembly of galaxies. In this context, it features a multiobject spectroscopy (MOS) mode covering a field of view (FOV) of $9~\mathrm{arcmin}^2$ and using 730$\times$342 individually addressable shutters for object selection. A variety of the JWST science goals also require the capability to conduct detailed spectroscopic studies of individual objects over a field of view of a few arc seconds. For that, NIRSpec is equipped with an integral field unit (IFU) with $3\times
3~\mathrm{arcsec}^2$ FOV, and five fixed slits for high-contrast
long-slit spectroscopy (SLIT mode). The instrument is sensitive
across the spectral range of $0.6\text{--}5.3~\mu$m.  This interval is
divided into three main scientific bands I, II, and III, which can be
selected by moving a matching long-pass filter into the optical beam.
In each band, two dedicated gratings provide a spectral resolution of
$R=\lambda/\Delta\lambda\approx1000$ and $R\approx2700$.  The complete
wavelength span can be observed in a single exposure with a
$\mathrm{CaF_2}$ prism at low resolution ($R\approx100$).  The nominal
instrument configurations, and corresponding combination of dispersive
elements and filters, are listed in \autoref{tab:NIRS_modes}.  For
target acquisition, NIRSpec can be put in imaging mode by selecting
the mirror instead of a disperser.

\begin{table*}
	\caption[]{Nominal NIRSpec operation configurations}
  	\label{tab:NIRS_modes}
	\centering
	\begin{tabular}{ccccc}
	\hline\hline
	Band & GWA element & Resolution $\lambda / \Delta\lambda$ & Filter & Spectral range / $\mu$m\\
	\hline
	I & G140M, G140H & 1000, 2700 & F100LP & 1.0--1.8 \\
	II & G235M, G235H & 1000, 2700 & F170LP & 1.7--3.1 \\
	III & G395M, G395H & 1000, 2700 & F290LP & 2.9--5.2 \\
	0.7 & G140M, G140H & 1000, 2700 & F070LP & 0.7--1.2 \\
	n/a & PRISM & ~100 & CLEAR & 0.6--5.3 \\
	IMA & MIRROR & n/a & F110W, F140X & 1.0--1.2, 0.8--2.0\\
	\hline
	\end{tabular}
	\tablefoot{Nominal combinations of filters and disperser
          elements in NIRSpec operations.  The gratings are used with
          long-pass filters, and the MIRROR is used with band-pass filters in
          imaging mode for target acquisition.  The additional filter
          position OPAQUE closes the instrument light path from
          outside and couples the internal calibration source
          CAA with the spectrograph.}
\end{table*}

In addition, NIRSpec features a suite of calibration sources both for
flat-fielding purposes and wavelength reference housed in the
calibration assembly (CAA) and coupled to the spectrograph via a
mirror on the back of the opaque filter. The instrument optical bench
and optical elements are mostly manufactured from silicon carbide
(SiC), a light and very stable material thermally. The instrument's
focal plane is equipped with two Teledyne ultra-low noise sensors
\citep{beletic+08}, provided by NASA Goddard Space Flight Center
(GSFC).

The complexity of the instrument, and in particular of the MOS-mode,
made it necessary already in the early stages of the project to
perform detailed calculations of the path the light follows entering
NIRSpec for simulation purposes and to assess the instrument
performance in detail. The NIRSpec parametric model was originally
developed to make predictive simulations of NIRSpec data and it
provides the basis of the NIRSpec Instrument Performance Simulator
(IPS); see \cite{Gnata:2007_00} and \cite{Piqueras:2008_00,
  Piqueras:2010_00}.  Naturally, this same model, once properly
adjusted to reflect the calibration data acquired during NIRSpec
cryo-vacuum test campaigns, also provides the transformations to
understand the origin of light falling onto the detectors, and as
such, it can be used i) to enable the extraction of wavelength
calibrated spectra and ii) to aid the required computations for the
onboard target acquisition procedure.  Most importantly, the ability
to extract wavelength calibrated spectra from any of the $\sim$250\,000 
slitlets using this model-based approach greatly improves the
efficiency with which data taken with the instrument can be
reduced. In particular, the conventional approach of carrying separate
empirical calibrations for each individual NIRSpec slitlet and each
disperser is clearly not practical, and with the model-based approach
the need to take matching calibration exposures during each
observation is no longer required.

As a consequence, the parametric model of the instrument became one of the core elements of the spectral and spatial calibration strategy of NIRSpec. The mission-level calibration requirements,
primarily reflecting the needs of redshift spectroscopic surveys, state that
the accuracy of the overall spectral calibration of NIRSpec has to be better than 1/4 of a pixel (root mean square; RMS). From this total budget, an allocation of 1/5 of a pixel has been set aside for the uncertainty of 
the parametric model and will be used to assess its accuracy. For the spatial calibration of NIRSpec, there are no mission-level requirements. However, the accurate extraction and registration of the spectra requires a model accuracy of 1/10 of a pixel for the NIRSpec spectrometer optics.

In this paper, we describe the process that we have developed to
adjust the NIRSpec parametric model on a dedicated set of calibration
data, and present a quantitative assessment of its ability to
reproduce the position and wavelength scale of the extracted
calibration spectra.
The calibration data used in this work were acquired during the
NIRSpec Performance Verification and Calibration campaign in
cryo-vacuum conditions, undertaken at the IABG testing
facilities in Germany in the first half of 2013
\citep{Birkmann:2012_00}. At the beginning of 2015, the Microshutter Assembly (MSA) and Focal
Plane Assembly (FPA)
units were replaced by newer, better performing units and the
instrument is only now in its flight configuration. Therefore, some of
the results presented here are not final and the parameters of the
instrument model will need to be updated to reflect the small changes
in the optical geometry of these elements, using data from a more recent
cryogenic test cycle. Nevertheless the same methodology described here
will be adopted for this step and for the on-orbit wavelength
calibration, and the same level of accuracy in the results presented
here is expected.

The paper is organized as follow. The concept and formalism of the
NIRSpec parametric model is introduced in Sect.~\ref{Model}.  In
Sect.~\ref{Data} the calibration data from ground testing used to
optimize the model are presented. The various data processing steps
and relevant software tools are described in Sect.~\ref{Processing}.
The procedure to optimize the model is described, together with its
intermediate results, in Sect.~\ref{Fitting}, while the final results
in terms of the parameter values and the spatial and spectral accuracy of the model
are presented in Sect.~\ref{Results}.  Finally, the reliability of our
results and prospects for the wavelength calibration of the instrument
once in space are discussed in Sect.~\ref{Discussion}.

\section{NIRSpec parametric model}\label{Model}

\begin{figure}[b]
   \centering
   \includegraphics[width=0.9\hsize, trim=0 3cm 0 0, clip=True]{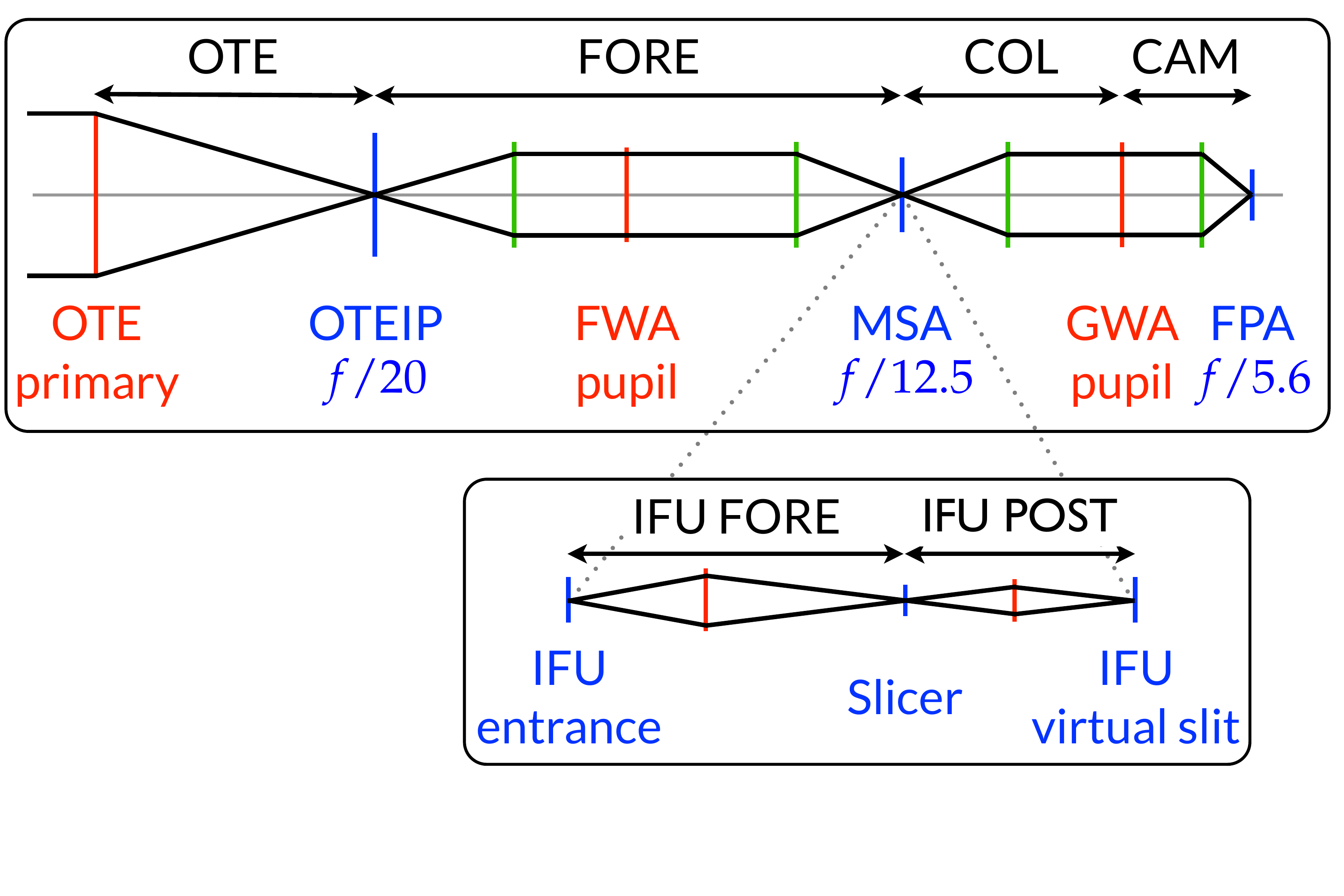}
   \caption{Paraxial layout of the JWST telescope and NIRSpec optical train with elements at principal planes and the insert for the IFU case. Focal planes are blue; pupil planes are red.}
   \label{fig:NIRS_paraxial}
\end{figure}

The science instruments onboard JWST are located behind the Optical
Telescope Element (OTE), which provides an image plane (OTEIP).  As
schematically shown in Fig.~\ref{fig:NIRS_paraxial}, the optical
design of NIRSpec consists of three major blocks, all employing
three-mirror anastigmats (TMAs) systems \citep{TePlate:2005_00}. The
light entering NIRSpec from the OTEIP is reimaged by the FORE optics
onto the aperture plane of the MSA.  The Filter
Wheel Assembly (FWA), which carries the filters for spectral band
selection (\autoref{tab:NIRS_modes}), sits in the pupil plane of the
FORE optics.
The collimator optics (COL) then projects the light from the slits
onto the Grating Wheel Assembly (GWA), where another pupil plane is located.
The GWA is equipped with various dispersers and a mirror for
imaging (see \autoref{tab:NIRS_modes}).  Finally, the camera optics
(CAM) focuses the (dispersed) beam onto the two detectors in the FPA.

The IFU entrance aperture is located in the MSA plane. For the majority of cases, IFU and MSA
observations are mutually exclusive as
their spectra share the same detector area and, therefore, all of the shutters have to be
closed during IFU operations and the IFU entrance has to be blocked
for multiobject exposures using the MSA.  The IFU optics are split
into an IFU FORE part, which reimages and rescales the MSA plane onto
the slicer, and an IFU POST part, which picks up the 30 image parts
and creates a virtual image for each slice at the MSA plane (virtual
slits).  The rest of the light path is then similar to the other
observation modes.

The NIRSpec parametric model encapsulates all of the main optical elements
identified in Fig.~\ref{fig:NIRS_paraxial}, but here we limit
ourselves to the optimization of the parameters describing the
spectrographic part of NIRSpec, that is the instrument from the MSA focal plane
to the FPA. Although a parametric
description of the PRISM is also part of the model, it was not optimized 
in this work and will be presented in a separate
paper (see also Sect.~\ref{Discussion}).

There are two types of components in the NIRSpec parametric model: the
parameterization of the coordinate transforms between the principal
optical planes (here COL, IFU FORE, IFU POST, and CAM) and the
geometrical description of the key plane elements (MSA, IFU slicer,
GWA, and FPA).  When we discuss coordinate transforms, we take the convention of naming so-called forward transforms
those following the direction of the light path in NIRspec (i.e., from
OTEIP to FPA). The type and number of the parameters specifying the
various elements of the parametric model are summarized in
\autoref{tab:NIRS_modelparams}. A more detailed description of
each element is given below.

\subsection{Optics}\label{sect:optics}

The TMAs in the NIRSpec optical train are manufactured to have only
small amounts of distortion.  It is therefore possible to model the
individual optical transforms with a paraxial transform between the
principal planes, and departures from the ideal paraxial system are
treated as distortions and modeled using a fifth-order, two-dimensional (2D) polynomial. The IFU-FORE
transform and the 30 IFU-POST transforms (one for each slice) also
employ this formalism.

The paraxial part of a forward coordinate transform is defined by the
magnification factors along the output axes $\gamma_x$ and $\gamma_y$,
the rotation angle of the coordinate system $\vartheta$, and the position
of the rotation center in the input and output reference frame
(x$_{0in}$, y$_{0in}$) and (x$_{0out}$, y$_{0out}$). Given a point
with coordinate (x$_{in}$, y$_{in}$) in the input plane, the equations
in the Appendix \ref{AppTrafoPlanes} provide
the formulas to compute its corresponding coordinate (x$_{out}$,
y$_{out}$) in the output plane, and vice versa.

\begin{table*}
	\begin{center}
	\caption[]{Summary of the parameters employed in the NIRSpec
          parametric model of the spectrograph. All of the parameters
          needed to transform a point from one focal or pupil plane
          are listed here. The two last columns
          give the total number of available and free parameters in the optimization process described in Sect.~\ref{Fitting}.}
  	\label{tab:NIRS_modelparams}
	\begin{tabular}{cccccc}
	\hline\hline
	 & & &\multicolumn{2}{r}{Number of Parameters}\\
	Element & Type & Description & Total & Free\\
	\hline
	 Optical module & Paraxial transform & Input plane center & 2 & \\
	 COL, CAM      & 		    & Output plane center & 2 & \\
	               & 		    & Rotation angle & 1 & \\
	               & 	            & Scaling factors & 2 & \\
		           & Geometrical distortion & Forward 2D polynomials & $21 \times 2$ & $21 \times 2$ \\
		       & 			& Backward 2D polynomials & $21 \times 2$ & $21\times 2$ \\
        \hline
	MSA            & Quadrant positions	& Position of shutter (1,1) & $2 \times 4$ & $2 \times 4$ \\
	               & 		    & Rotation angle & 4 & 4 \\
	               & Shutters            & Pitch size & $2 \times 4$ & $2 \times 4$ \\
	               & 	            & Aperture size & $2 \times 4$ & \\
	               & Fixed slits         & Absolute position & $2 \times 5$ & $2 \times 5$ \\
                       &                    & Aperture size & $2 \times 5$ & \\
        \hline
	GWA            & Dispersers & Alignment angles & $3 \times 8$ & $3 \times 6$ $\dagger$ \\
	               & Grating dispersion & Front surface tilt & 6 & \\
	               & 		    & Groove density & 6 & \\
	               & PRISM dispersion   & Front surface tilt & 1 & \\
	               & 		    & Internal prism angle & 1 & \\
	               & 		    & Sellmeier refractive index & 8 & \\
	               &				& Temperature and \\
	               &				& pressure dependence & 6 & \\
        \hline
	FPA            & Detector positions & Position of pixel (1,1) & $2 \times 2$ & 2 $\ddagger$ \\
	     	       &		    & Rotation angle & 2 & 1 $\ddagger$ \\
	               & Pixel positions    & Pitch size & $2 \times 2$ & \\
        \hline
	IFU            & Slicer position & Absolute position & 2 & \\
	               & 				  & Rotation angle & 1 & \\
	               & Slices           & Relative position & $2 \times 30$ & \\
	               & 				  & Aperture size & $2 \times 30$ & \\
IFU-FORE       & Optical module  &  Paraxial + 2D polynomials & $7 + 21 \times 2$ & \\
 	IFU-POST       & 30 Opt. modules &  Paraxial + 2D polynomials & $(7 + 21 \times 2) \times 30$ & $3 \times 30$ $\clubsuit$ \\
	\hline
    \hline
	\end{tabular}
\end{center}

$\dagger$ The prism was not included in this work and the mirror alignment was set to fixed angles.\\ 
$\ddagger$ Only for SCA492.\\
$\clubsuit$ Only the center position and rotation of the output plane of each IFU-POST paraxial part were  optimized.\\

\end{table*}

\subsection{Microshutter Assembly} 

\begin{figure}[b]
   \centering
   \includegraphics[width=0.9\hsize, trim=5cm 1.4cm 3cm 1.5cm, clip=True]{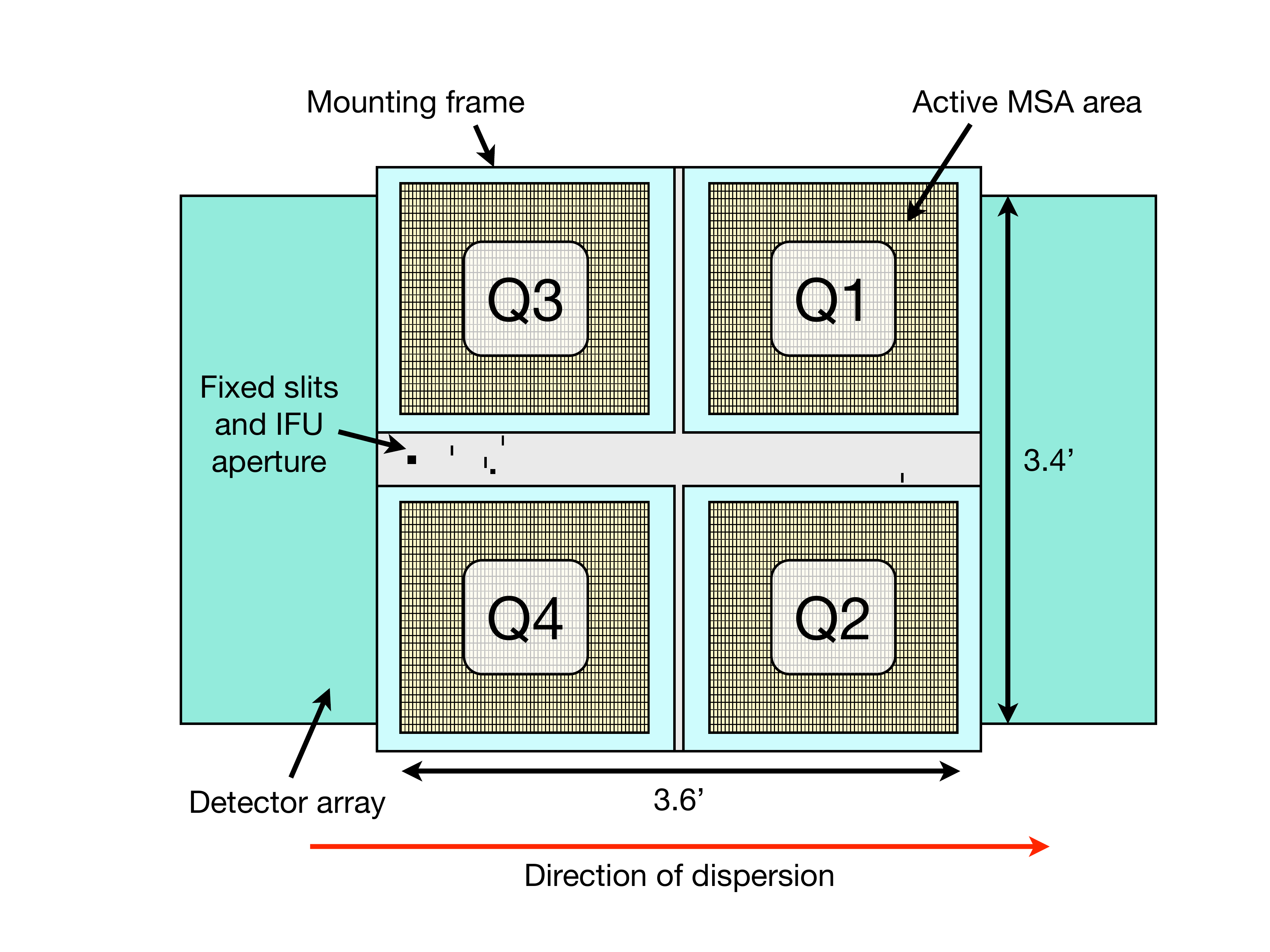}
   \caption{Geometrical layout of the MSA projected onto the detector
     plane. The quadrants Q1--4 contain the microshutters;  5 high-contrast slits and IFU aperture positioned in the horizontal area in the center 
     are indicated.}
   \label{fig:MSA_geo}
\end{figure}

The MSA is located at the slit plane of the instrument's
spectrographic part. This element, built at NASA GSFC, features four
arrays of 365$\times$171 microshutters, each
0.2\arcsec\,$\times$ 0.46\arcsec\ in size once projected onto the
sky (respectively in the dispersion and cross-dispersion directions).
The microshutters are arranged in a 2$\times$2 mosaic, covering in total a FOV of
$9~\mathrm{arcmin}^2$ with $\sim$250\,000 individually addressable
slitlets \citep{Kutyrev:2008_00}. Mechanically, the microshutters have been
implemented using micro electromechanical switches (MEMS), which each have
an aperture size of 76 $\mu$m $\times$ 175 $\mu$m. The geometrical layout of the MSA module is shown in
Fig.~\ref{fig:MSA_geo}. As depicted in the figure, between the top
and bottom quadrants, the five fixed slits are positioned together
with the IFU aperture. Three of the high-contrast long slits have
widths of 0.2\arcsec, and one is 0.4\arcsec\ wide, and one is a square
aperture of 1.6\arcsec\,$\times$ 1.6\arcsec. The reference frame
$x$-axis is aligned with the dispersion direction.

In the parametric model, the geometry of the MSA element is captured by
a total of 40 parameters. For each of the four microshutter
quadrants, we use the 2D coordinates of shutter $(1,1)$ relative to the center of the
FOV, a rotation angle, plus the shutter pitch size in $x$ and $y$ (for a total
of 20 parameters).  For each fixed slit, similarly, we have the
center position and size in both dimensions.  Each microshutter is
labeled with an index for the quadrant and two indices denoting the column and row,
$(q, i, j)$, while the fixed-slits have names (S200A1, S200A2, S200B1,
S400A1, and S1600A1). The MSA model parameters are used to transform 
from a shutter array index or slit ID with a simple geometrical calculation
to its absolute physical position in the MSA plane (corresponding to the COL
input plane), and vice versa.

\subsection{Grating Wheel Assembly}

The NIRSpec GWA is a cryogenic wheel
mechanism that can be configured to position one of its optical
elements into the beam path. It is equipped with six dispersion
gratings ($R\approx1000$ and $R\approx2700$), a prism ($R\approx100$),
and a mirror for the imaging mode (\autoref{tab:NIRS_modes}). The
rotational degree of freedom of the wheel is given by a ball bearing
controlled by two mechanisms: a cryogenic torque motor used as
actuator and a spring operated ratchet to achieve accurate
positioning.  Despite the excellent mechanical performances of the
grating wheel, the finite reproducibility of the positioning still
causes small random displacements of the light beam on the focal plane
each time the wheel comes back to a given position. To be able to
predict the position of the light beam, two position sensors are used
to accurately measure the tip and tilt displacement of each selected GWA
element \citep{wfe+2008}.

In the parametric model, the orientation of the GWA coordinate frame
is such that the $x$-axis points in the dispersion direction, $y$ in cross
dispersion, and $z$ in the beam direction. The diffraction of the six
gratings is specified by their individual groove densities and front
surface tilt angles.  The double-pass PRISM dispersion is characterized by the
front surface tilt, internal prism angle, and a relation  with wavelength, temperature, and pressure
dependence for the
refractive index.  The MIRROR is treated as a simple reflective surface.

The orientation of each disperser with respect to a reference surface
at the GWA is given by four angles: one macroscopic front surface tilt
angle between the grating surface and reference surface,
$\Theta_y$, plus three individual alignment angles $(\vartheta_x,
\vartheta_y, \vartheta_z)$. The formalism that we employ to transform
a light ray coordinate in the output plane of COL to the input plane
of CAM passing through one of the elements of GWA are given by
Eq.~\ref{eq:rots1} to Eq.~\ref{eq:p5} in the Appendix \ref{AppGWATransforms}.
 
\subsection{Focal Plane Assembly}

NIRSpec's FPA is equipped with two 5.3 $\mu$m cutoff, Teledyne
HAWAII-2RG sensor chip assemblies (SCAs), provided by NASA GSFC
\citep{rab+08, Rauscher+2014}. These HgCdTe sensors are ultra-low 
noise, state-of-the-art, near-IR detectors, which each have 2048$\times$2048 
pixels with 18 $\mu$m pitch size (corresponding to 100 mas on
the sky); they are labeled SCA491 and SCA492, and aligned along the
dispersion direction of the focal plane with SCA491 on the blue side
(short wavelengths). The dispersion direction defines the $x$-axis of
the FPA.

In the parametric model, the geometry of the FPA is captured by ten
parameters: the absolute position 2D-coordinates, a rotation angle, and
pixel-pitch in $x$ and $y$ for each array. These parameters are used
to transform with a simple geometrical calculation from a pixel array index
to its absolute physical position in the FPA plane (corresponding to the CAM output
plane), and vice versa.

\begin{table*}
	\begin{center}
	\caption[]{Information summary of the type of exposures
          acquired during NIRSpec testing, which were then used to
          extract the different type of reference data necessary for the model optimization. FXSL indicates the fixed-slits.}
  	\label{tab:NIRS_refdata2}
	\begin{tabular}{ccccc}
	\hline\hline
	Source Type (Name) & Slits & GWA & Reference data  & Reference type\\
	\hline
	Continuum (TEST) & MSA all open, FXSL, IFU open & MIRROR & MSA quadrant positions, & Spatial\\
	 &  &  & SLIT and IFU positions & \\
	Continuum (TEST) & MSA checkerboard, FXSL & MIRROR & MSA shutter positions, & Spatial\\
	 &  &  & SLIT positions & \\
	Continuum (FLAT1,2,3) & 4 MSA dashed-slit patterns, FXSL & All gratings & Spectrum locations & Spatial\\
        Continuum (FLAT3) & MSA closed, IFU open &  G395H & Spectrum locations & Spatial\\
	                  & MSA closed, IFU closed & G395H &  Failed open shutter spectra & $\dagger$   \\
	Absorption lines (REF) &  FXSL & All gratings & Spectrum locations, & Spatial\\
	                &      &          &  spectral feature locations &  Spectral\\
	Emission lines (Argon) & 4 MSA dashed-slit patterns, FXSL & All gratings & Line positions & Spectral\\
			      & MSA closed, IFU open & G395H &  Line positions & Spectral  \\
			      & MSA closed, IFU closed & G395H &  Failed open shutter spectra & $\dagger$ \\
	\hline
	\end{tabular}
	\end{center}
$\dagger$ MSA-background exposure for the IFU exposure
\end{table*}

\subsection{Integral field unit}

The NIRSpec IFU design is based around an image slicer element
consisting of 30 stacked mirror surfaces that are curved and tilted with
respect to each other so that the image is split into 30 individual
slitlets, each of which is directed onto a dedicated pupil mirror
\citep{Closs:2008_00}. In this way, the 3\arcsec\,$\times$ 3\arcsec\ square FOV is
dissected into 30 slices of $0.1$\arcsec\ width (spectral direction) and
$3$\arcsec\ length (spatial direction). The spatial direction of the virtual
slit is sampled at the FPA by 30 $0.1$\arcsec-pixels, resulting in an IFU
field consisting of 30$\times$30 spaxels, each of size 0.1\arcsec\,$\times$
0.1\arcsec.

In the parametric model, the optical modules in the IFU are captured
with the same combination of a paraxial transform with superimposed
polynomial distortions as all of the other NIRSpec optics described above
(Sect.~\ref{sect:optics}).  The IFU FORE optics consists of a
transform between the MSA and slicer focal plane.  The IFU POST part
comprises 30 individual coordinate transforms, one for each slice,
from the slicer to the virtual slit plane at the MSA.  The IFU slicer
itself is described by its absolute position and a rotation angle (in
the IFU FORE output plane), whereas each slice is described by a
relative position in the slicer and aperture size.

%__________________________________________________________________
\section{Calibration data}\label{Data}

The data used for the model optimization process were acquired during the
NIRSpec flight model performance verification and calibration campaign
undertaken at the IABG testing facilities in Ottobrunn (Germany) in
2013. During two cycles of cryo-vacuum
testing, NIRSpec was placed into a cryogenic chamber, under an inner
shroud cooled by liquid Helium to $\sim$40 K, corresponding to
NIRSpec operating temperature once in space.  During testing, data
were acquired from various light sources. We used both internal
sources (housed in the CAA) and external lamps, housed within the
cryogenic test setup and coupled to NIRSpec via the instrument
pick-off mirror. All of the sources discussed here provide a spatially uniform
illumination of the NIRSpec slit plane \citep{Birkmann:2012_00}.

Three types of
data are necessary for the optimization of the NIRSpec parametric model: imaging data, continuum spectra, and
spectral lines. The first two sets provide spatial reference data,
while the last set is used to derive spectral references.  Illumination
for imaging mode is provided by the CAA lamp labeled TEST; for
continuum we used the CAA sources FLAT1, FLAT2, and FLAT3 that
provide blackbody spectra for the three wavelength ranges
$1.0\text{--}1.8~\mathrm{\mu m}$, $1.7\text{--}3.0~\mathrm{\mu m}$, and
$2.9\text{--}5.0~\mathrm{\mu m}$, matched to the NIRSpec bands
(\autoref{tab:NIRS_modes}).  The spectral references are provided by
the CAA source labeled REF, which is a rare earth absorption line source
(Erbium) with absorption features in the $1.4\text{--}1.6~\mathrm{\mu m}$
wavelength range, and an external Argon source, which emits
unresolved lines in the range up to 2.5 $\mu$m. For these sources, the gratings of
Band II and Band III were used in second and third order (as opposed
to the default first order as in Band I), while the FORE band-pass filters removed 
overlaps of the higher orders.

The data were acquired
by cycling through exposures with the relevant light sources and different MSA configurations for each selection of disperser or for the mirror. To prevent overlapping among spectra
when using any of the gratings, only one microshutter per MSA row was
opened \citep{Ferruit:2012_00}.  The type of data collected for
the model optimization are summarized in \autoref{tab:NIRS_refdata2}
and described below in further detail.

\subsection{Spatial references}
\label{sec:spatial_refs}

\begin{figure*}
	\centering
	\includegraphics[width=\textwidth, page=1, clip=true, trim=0 5cm 0 5cm]{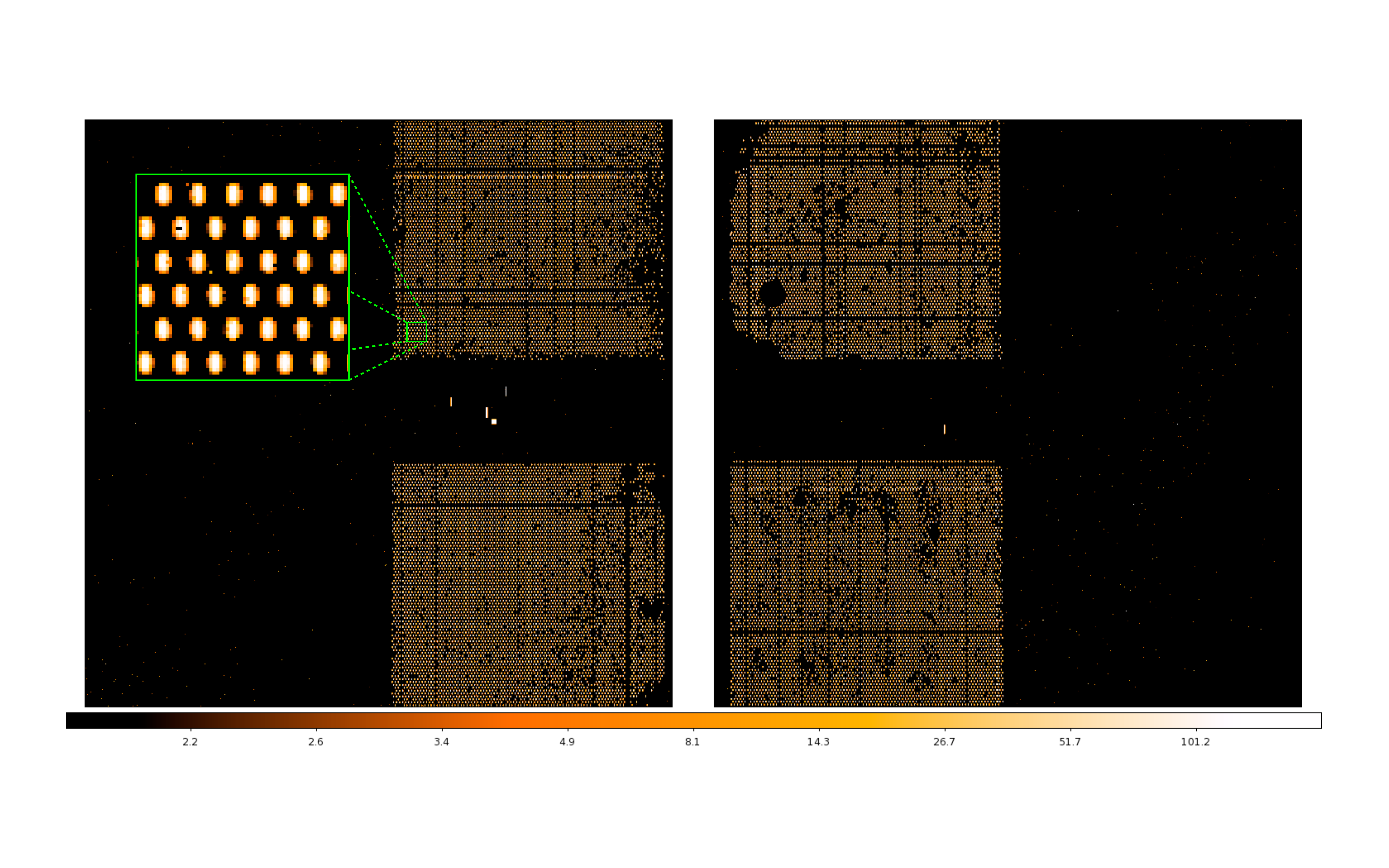}
	\caption{\label{fig:msa_checkboard} Count-rate image of the two
          NIRSpec detector arrays (SCA491 and SCA492) of an
          exposure of the TEST lamp acquired in imaging mode (MIRROR)
          with the MSA configured in a checkerboard pattern, more clearly visible in the enlargement.
          SCA491 is on the left (blue side) and SCA492 on the right
          (red side).  The images of the fixed slits are in the center
          of the images. Areas of defective shutters that fail to open
          are clearly visible and, in some cases, a full row or column is
          fully closed. The dark round feature in SCA492 is an area of
          dead pixels.  This image was acquired with the old set of detectors and old MSA module. The image was generated from NIRSpec raw
          data following the steps described in Sect.\,\ref{Processing}.}
\end{figure*}

For the imaging mode, we acquired exposures of the MSA, illuminated by
the TEST lamp, in two different configurations: {\em i)} commanded
fully open and {\em ii)} in a checkerboard pattern, a regular
pattern of one open shutter for every three closed shutters, which can be seen in the enlargement in
Fig.~\ref{fig:msa_checkboard}. In the full image, one can clearly notice the
imprint of the shutters that fail to open, either because they are defective or
because an entire row or column was masked to prevent electric
shorts in the MSA arrays, in which case no shutter in that line can be
commanded open. 

When configured to fully open, the IFU aperture was also open,
providing an image of the 30 virtual slits. The fixed slits are always
open. As described in Sect.~\ref{Fitting}, these imaging exposures
are used to first determine the approximate location of the MSA
quadrants projected to the detector (fully open configuration) and then to derive the accurate
FPA coordinates of the open shutters (checkerboard exposure).

To determine the FPA position of the spectra in the spatial direction, for
each grating selection we acquired four exposures, with the MSA
configured in four different patterns of multiple so-called dashed-slits and
illuminated by the continuum flat-field lamp appropriate for that
grating (i.e., FLAT1 for G140M and G140H, FLAT2 for G235M and G235H, or
FLAT3 for G395M and G395H). A so-called dashed-slit consists of 14 shutters 
in a column, where an open shutter is followed by a closed one,
yielding seven open shutters per slit. The four patterns
were chosen to properly sample the MOS FOV, and are identical to those chosen
for the acquisition of the spectral reference data. We placed ten dashed slits distributed in each
quadrant, using alternatively Q1 and Q2, or Q3 and Q4.
Fig.~\ref{fig:dashed_lowres} provides an image of spectra taken
with one of the four MSA dashed-slit patterns.

\subsection{Spectral references}

\begin{figure*}
	\centering
	\includegraphics[width=\textwidth, page=2, clip=true, trim=0 5cm 0 5cm]{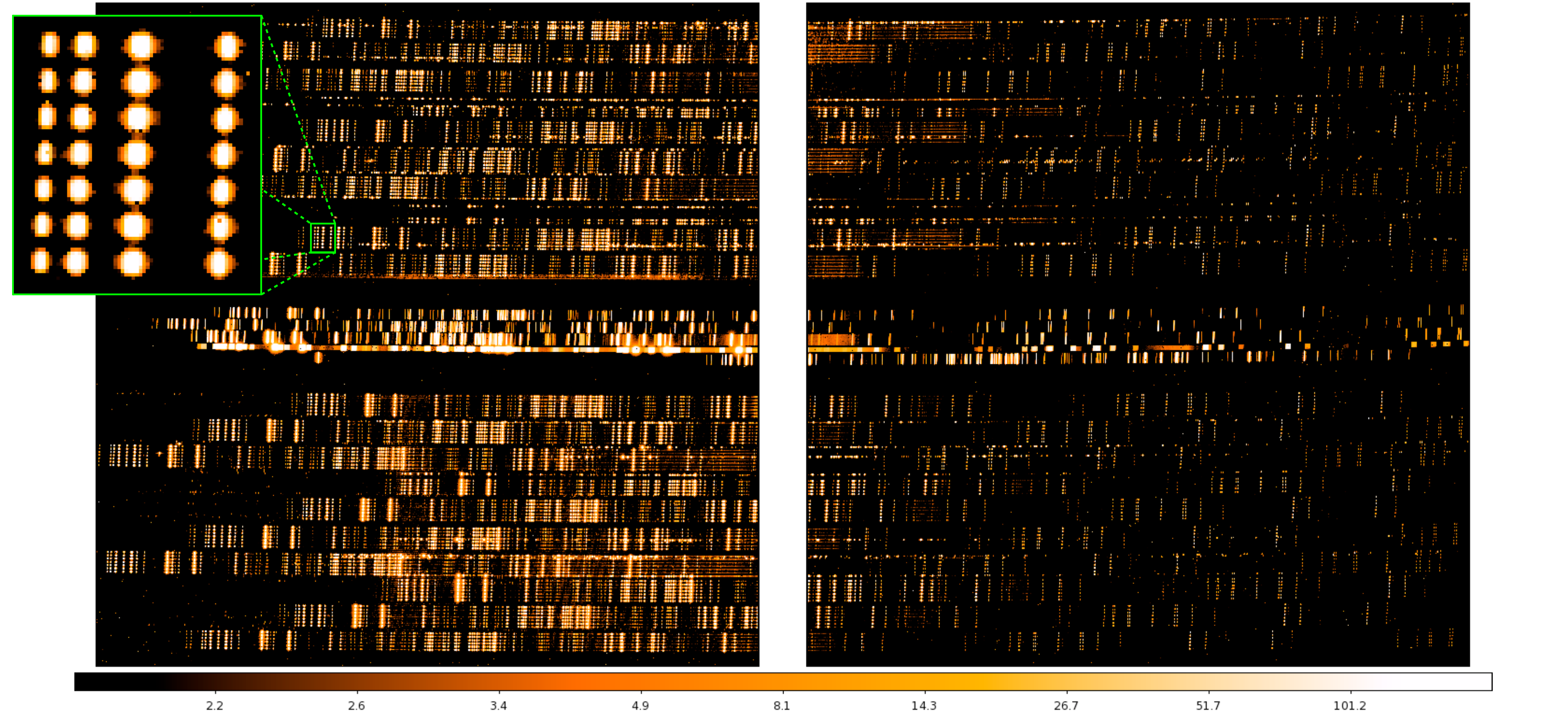}
	\caption{\label{fig:dashed_lowres} Count-rate image of the two NIRSpec
          detector arrays (SCA491 and SCA492) of an exposure of
          the Argon lamp acquired with G140M and the MSA configured in
          a dashed slit pattern (see enlargement). In this case the dashed slits were
          open in Q3 and Q4. The apparent width of the lines depends on the image scaling 
          and blending of unresolved pairs. The regular pattern of spectra is
          spoiled by the missing spectra due to shutters that fail
          to open, or by misaligned, overlapping spectra from shutters
          that cannot be closed (stuck open).
        The image was generated from NIRSpec raw data using our preprocessing pipeline.}
\end{figure*}

 Four exposures of the REF source were acquired for each
grating for a first approximate measure of the location of the spectra on the
FPA. For more accurate measurements of the FPA positions of a set
of reference lines, we used four exposures of the Argon source for
each grating with the MSA configured in the four dashed-slit patterns
described in Sect.\,\ref{sec:spatial_refs}. An example of these
exposures for G140M can be seen in Fig.~\ref{fig:dashed_lowres}. In
this image one can notice the missing spectra from the failed
closed shutters, as well as misaligned overlapping spectra from
defective shutters that cannot be commanded closed (stuck open).

The Argon source was also used in combination with a fully closed
MSA pattern and open IFU aperture and the grating
G395H. To deal with the problem of failed open shutters interfering with
the Argon spectra from the IFU, we also acquired Argon lamp exposures
with a fully closed MSA and closed IFU aperture,  providing
an image of the contaminating signal of the failed shutters, which then can
be subtracted away.

%__________________________________________________________________
\section{Data processing}\label{Processing}

All of the data presented here were acquired in full frame up-the-ramp mode, in which each detector pixel is sampled nondestructively every 10.74 s (see \citealt{Rauscher:2007_00} for more details on the instrument readout scheme).
The two raw data cubes from all of the exposures were first processed to derive the count-rate images.
This involves the following steps: bias subtraction, reference pixel subtraction, linearity correction, dark subtraction,\footnote{Dark subtraction is carried out at the data cube level, i.e., the corresponding frame from a low-noise dark-currents cube is subtracted from each frame of the exposure.} and up-the ramp fitting, as described in more detail in \cite{Birkmann:2011_01} and \cite{Boeker:2012_00}.
The output of these preprocessing steps are two \verb+FITS+ files (one for each detector) with image extensions for count-rate, variance, and quality flags for each pixel. Examples of count-rate images are shown in Fig.~\ref{fig:msa_checkboard} and \ref{fig:dashed_lowres}. If necessary, count-rate images of dedicated  background exposures were subtracted from the calibration images to remove the spurious signal of failed open shutters.

When dealing with imaging exposures, preprocessing is all that is
needed before analyzing the data.  However, further computation steps
are necessary for the spectral data.  To extract spectra of the
NIRSpec fixed-slits, MSA microshutters or the IFU from the count rate
images, we developed the dedicated NIRSpec IPS Pipeline Software
(NIPS), which uses the parametric model described in Sect.\ \ref{Model}
to perform the following operations.

After having initialized the instrument parametric model to the appropriate observing mode, the location of the spectrum trace for a given slit, i.e., the curve of the slit center projected to the detector, is calculated between the lower and upper boundaries of the mode wavelength range using the model forward transforms for the slit coordinates in the MSA plane.
The detector areas are then extracted as a subimage.
Wavelength $\lambda$ and spatial coordinates $d_y$ are assigned to each pixel in the subimages according to the combination of slit, filter, and grating by a computational method that we call ``meeting at the grating'', described in Appendix \ref{AppMeeting}. 
The result of this operation is a 2D spectrum, irregularly sampled in $\lambda$ and $d_y$.
From this data, it is already possible to calculate irregularly sampled spectra with extraction operations working in the pixel columns or pixels belonging to wavelength bins.

The next step is the generation of a so-called 2D-rectified spectrum, for which the data in the subimage is resampled onto a regular grid of wavelength and spatial coordinates.
Finally, the one-dimensional (1D) rectified spectrum is computed as the median  along the spatial direction in each wavelength bin of the 2D-rectified spectrum using only pixels with the selected quality flags.
For all of the data sets, dead (i.e., unresponsive) pixels were flagged and not used in the computation of the 1D spectra.
The spectra were all derived at the NIPS default resolution for each disperser, given by the Nyquist sampling of the nominal resolution of the disperser at the central wavelength (\autoref{tab:NIRS_modes}).

The NIPS only utilizes a single thread per execution, which means that the extraction and rectification of hundreds of spectra for each grating was a lengthy process; in fact, in terms of computation time this was the lengthiest step of the model calibration.
On average, the extraction and rectification of 100 MOS spectra from a high resolution grating took about 30 minutes on a state-of-the-art workstation CPU.

\subsection{Reference data}\label{refdata}

After these general processing steps, the reference data used by the model optimization procedure were generated.
For the imaging mode, we measured the centroids of the slit images.
The input was an exposure with the 3$\times$3 checkerboard pattern (similar to Fig.~\ref{fig:msa_checkboard}) during which the fixed slits and IFU virtual slits were also illuminated.
At first, the location of each open shutter on the detector was predicted with our initial guess for the parametric model.
The shutter images typically cover 2$\times$5 pixels, and their surrounding area was cut and examined for bad pixels.
If this subimage contained invalid pixels it was discarded, along with those of known failed closed shutters or those with insufficient signal (because the corresponding microshutter had unpredictably failed to fully open).
Finally, the geometrical centroid of the area around the predicted slit image was calculated, while taking care not to include pixels illuminated by the adjacent open shutters.
The subimages of fixed slits and IFU virtual slits were processed in a similar way, but were not filtered for defective pixels since the slit images are much larger in the spatial direction and, therefore, their centroids are less influenced by bad pixels.

The reference data from spectra were created with the NIPS using an intermediate model in the course of iteratively adjusting the parameters.
The final measurement output for a slit consisted of pixel coordinates $(i_{ref}, j_{ref})$ as a function of the true wavelength $\lambda_{ref}$.
To obtain accurate reference points despite a model with, at this point, still insufficient accuracy, the following approach was taken.
For the references in the spatial direction, the location of the continuum spectra for the opened shutters were predicted with the available model and subimages were cut from the data.
To measure the spectral trace position, the centroid in each pixel column was calculated in a range slightly larger than the spectrum width.
The results were restricted to areas without contamination by neighboring or failed open shutters.
For each spectrum, the centroid along the columns was then fitted with a fourth-order polynomial $j=P_j(i)$ with $4\sigma$ rejection of outliers, in particular to eliminate the small-scale influence of bad pixels.

For the references in the spectral direction, the Argon line spectra were rectified and collapsed to 1D using the median value across the slit.
For each grating, a set of 12--35 isolated lines was selected from the spectrum and the profiles were individually fitted with a Gaussian in the range of $\pm 5$ spectral pixels around the predicted line location.
For a given line, this fit provided the wavelength assigned by the model $\lambda_m$.
The model-predicted detector pixel position of the line $(i_m, j_m)$ was then calculated with the coordinate transforms from the slit to the detector for $\lambda_m$.
However, the true detector position $(i_{ref}, j_{ref})$ is different, since the intermediate model was not perfectly accurate.
In the spatial direction, the model trace generally does not match the measured trace $P_j(i)$.
As a result of the line profile fit and model calculation, $i_m$ is the coordinate of the intersection between the model trace and line image.
Because of the slit tilt and the trace offset, $i_m$ is therefore different from $i_{ref}$.
Typical values for the trace position difference of the gratings were about 0.5 px (see below in \autoref{tab:resopti}), and the slit tilt is <8\degr\ for the R2700 and <4\degr\ for the R1000 gratings.
Therefore, the difference in the spectral direction coordinates was generally <1/14--1/29 of a pixel, which is far below the 0.2 px RMS allocation in the spectral calibration budget, and was neglected by setting $i_{ref} = i_{m}$.
The reference position in the spatial direction was then computed from the trace polynomial as $j_{ref}=P_j (i_m)$.
Combining the pixel coordinates and the true wavelength of the emission line, the reference tuple $(i_{ref}, j_{ref})(\mathit{slitID}, \lambda_{ref})$ was obtained for each slit and line and could then be compared to the model calculated positions for the true wavelength $(i_m',j_m')(\mathit{slitID}, \lambda_{ref})$.

%__________________________________________________________________
\section{Model optimization procedure}\label{Fitting}

Because NIPS uses the transforms of the parametric model to obtain the reference data, the optimization of the model had to follow an iterative approach.
The spectra from the different reference data sets listed in \autoref{tab:NIRS_refdata2} were extracted from the count-rate images with an increasing level of accuracy in terms of wavelength and spatial location within the slit, while we progressed through the various steps of our model optimization process.
At first, NIPS operated with our initial best-guess of the NIRSpec parametric model and the output data are characterized by only approximate wavelength and spatial coordinates.
The model parameters were then adjusted, manually at first, and then via a proper optimization procedure, so that the final results reach a high level of accuracy.

The initial parameters of the instrument model were obtained from subsystem tests or optical modeling.
The manufacturer Reosc (Sagem) delivered a parametric description of the TMA modules to Airbus DS, who then assembled a complete as-built optical model of NIRSpec in the commercial optical design software Zemax\footnote{\url{http://zemax.com/}}.
We traced grids of rays through this prescription and derived the parameters of the paraxial and distortion transforms from the calculated ray coordinates.
The IFU transforms were obtained in a similar way, but here we relied on the as-designed Zemax model of this subsystem.

The information about the fixed slits and MSA quadrant positions and pitches are from subsystem measurements at warm transformed to cold conditions and were supplied by NASA GSFC.
The grating dispersion and front surface tilts were set to the ambient design values provided by the manufacturer.
We had no information about the relative alignment angles of the GWA elements, so they were set to 0 in the beginning.
As a starting point for the FPA, we assumed perfectly aligned detectors without rotation in accordance with the nominal values provided by NASA GSFC.

\subsection{Model assumptions}\label{Assumptions}

The instrument description partly offers more flexibility than needed, since some parameters are degenerate.
For instance the distortion polynomials have low-order terms that mimic a magnification or constant offset, which are also present in the paraxial description.
Then again, such offsets can be captured by the parameters of the focal plane elements.
To avoid degeneracies, set absolute references, and avoid unnecessary complexity, the final model was created with the following premises:

\begin{itemize}
	\item The GWA MIRROR has all alignment tilt angles at 0 and defines the GWA reference plane. All other dispersers have alignment tilt angles relative to this surface.
	\item The MSA quadrants are rectangular and regular in size, i.e., in an individual quadrant the shutter pitch is uniform and the shutter axes are perpendicular.
	\item The FPA gap is forced to be centered on the y-axis. SCA 491 has no rotation and is symmetrical to the x-axis. SCA 492 is free to move and rotate within the first condition that couples the positions of 491 and 492. This does not restrict the modeling of the instrument, as the movements can be compensated by the distortion polynomials of the CAM. It does, however, produce a geometrically simple FPA description without excessive tilts and offsets.
	\item The COL and CAM transform are assumed to be achromatic (all-reflective optical parts) with no wavelength dependence of the distortion coefficients.
\end{itemize}

\subsection{Manual adjustment}

As mentioned before, the model fit was carried out based on data extracted with an unoptimized initial model, hence the pixel coordinates of the spectra were not accurately known.
To avoid the contamination of reference data by nearby slits and shutters during the extraction, it was necessary to locate the spectra with an initial accuracy better than 1 pixel, in particular in the spatial direction.
Therefore, the first steps in the calibration were a manual adjustment of the detector, fixed slit, and MSA quadrant positions and pitches.
Imaging data of the slits and the fully open MSA were plotted with the projected slit and MSA shutter outlines.
Known failed closed shutters created a quasi-random pattern, in particular at the quadrant edges, and allowed us to match the exposure with the model prediction.
From the offsets, the change of the positional parameters was visually deduced and fed into the model.
The same approach was taken to narrow down the positions and rotations of the IFU virtual slits, although only after having optimized the spectrograph optics.
In addition, spectra of the fixed slits with the Erbium absorption source were used to estimate an offset in the spectral and spatial direction.
In this case, the alignment angles of the dispersers were adjusted to achieve a visual match between the predicted and measured positions of the traces and known absorption features.
The intermediate model was then accurate enough to enable the derivation of the reference data as previously described in Sect. \ref{Processing}.

All of the necessary processing and visualization scripts for the manual adjustment had been prepared and tested with simulated data beforehand.
Therefore, this step could be carried out very efficiently once the real data were available and it was completed within a day.

\subsection{Automated optimization}

\begin{figure}
	\centering
	\includegraphics[width=\columnwidth, trim=0.65cm 0cm 0 1cm, clip=true]{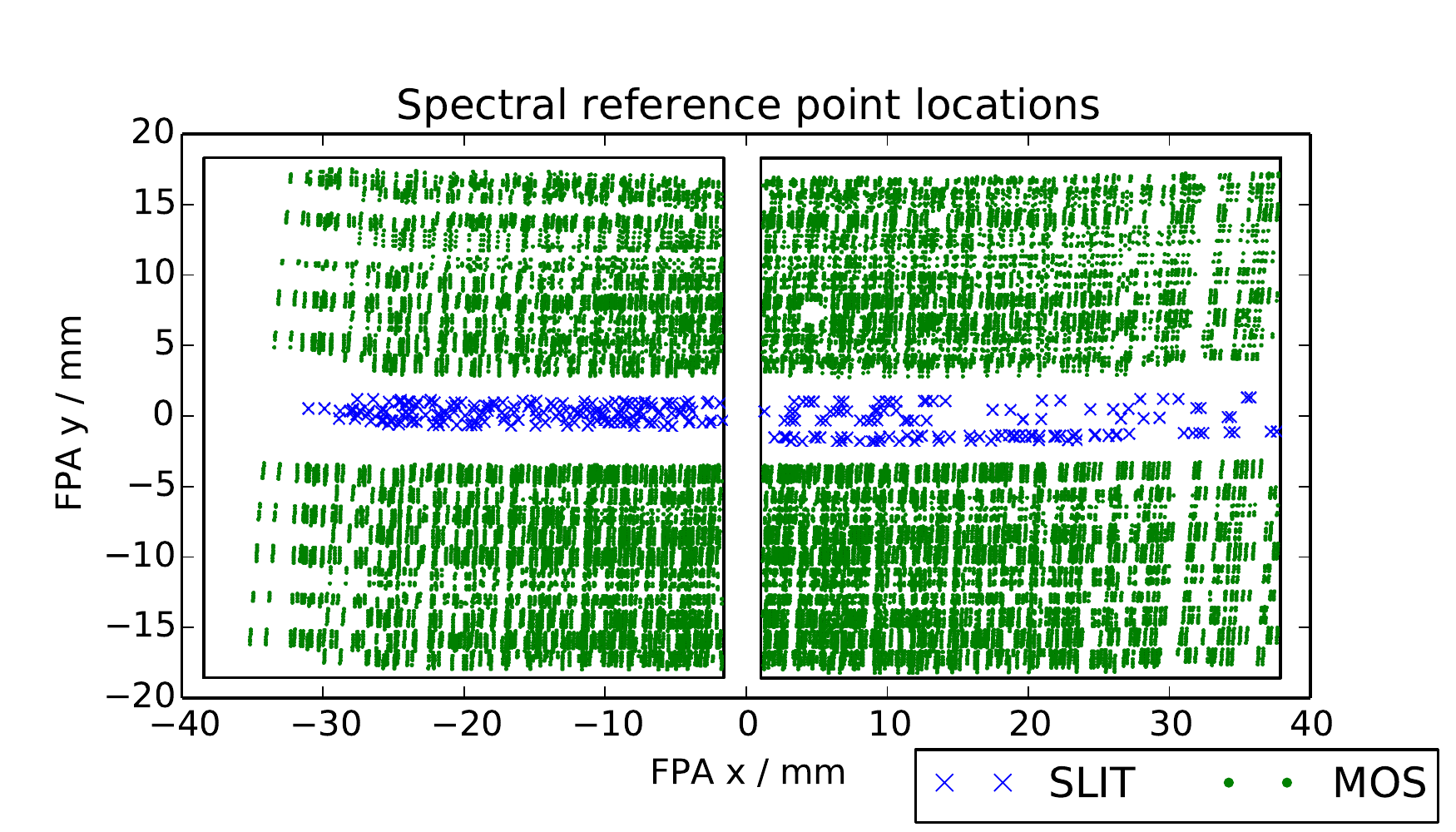}
	\caption{\label{fig:refspectral} Location of spectral reference points in the global model optimization. These reference points consist of the positions of Argon emission lines taken with the gratings G140M, G235M, G395M, and G395H. In total, there are 28\,020 points distributed over both detectors. The top and bottom groups of green dots are from shutter spectra; the central blue crosses from the fixed slits.}
\end{figure}

The calibration of the parametric model was eventually carried out with an automated optimization in multiple steps.
The common principle was to minimize the residuals between model predicted and true coordinates from the reference data with least-squares fits, during which selected subsets of the parameters were modified.

The first step of the optimization covered the forward transform from MSA to the detector.
In this step, we treated the global characteristics of the spectrograph, hence 120 parameters were fitted simultaneously, consisting of MSA quadrant positions, pitches and rotation, COL forward distortion, GWA disperser tilt angles, CAM forward distortion, and FPA positions and rotation (compare with \autoref{tab:NIRS_modelparams}).
The fixed slit positions in the MSA were not optimized at this stage because their images in combination with the predefined MIRROR orientation served as an absolute reference for the detectors.
To obtain a reliable and spatially unbiased description of the optical distortion, the reference data set was constructed from a selection of the GWA elements whose points cover the FPA plane as evenly and wide as possible.
The gratings G140H and G235H have an asymmetric distribution of lines on the detectors concentrated on SCA 492, and these were left out in the beginning to prevent an imbalance in favor of this side of the field.
Therefore only the data of the gratings G140M, G235M, G395M, and G395H were used, which each have 4000--10\,000 reference points that extended almost fully up to the FPA edges and evenly cover the two SCAs.
Combining the data of the fixed slits and four different MSA configurations per grating led to a total number of 28\,020 spectral reference points as shown in Fig.~\ref{fig:refspectral}.
The MSA imaging reference data contained 23\,727 centroids in total.
To obtain a similar amount of points as for the other GWA elements, about 1500 of them were randomly selected per quadrant, yielding 5982 reference centroids, including the fixed slits.
Therefore, in total 34002 points were available for the global model optimization.
We defined a function comparing the pixel positions calculated by the model with the reference data, and we minimized the residuals by optimizing the model parameters with a least-squares fit.
To reject outliers, the points were filtered with a $4\sigma$ clipping in each subset of the same slit type (fixed slit or MOS), pixel coordinate direction, GWA element, and detector in each residual calculation.
Despite the large amount of reference points and free parameters, the least-squares minimization process was not particularly computationally expensive and was run on a normal workstation, taking approximately 1 hour (using only one thread).

\begin{table*}
	\caption{Residuals of the forward projection from MSA to FPA before and after the global model optimization from gratings and the MIRROR on the detector.}
	\label{tab:resopti}
	\centering
	\begin{tabular}{ccccccc}
	\hline\hline
	GWA		& Model	& \multicolumn{5}{c}{Residual / pixels}\\
	elements& status& $i$ mean + RMS & $i$ median (68.27\%) & $j$ mean + RMS & $j$ median (68.27\%) & 68.27\% absolute \\
	\hline\\[-0.9em]
	Gratings & initial	& $-1.757\pm1.654$ & $-2.212^{+2.742}_{-1.221}$ & $0.534\pm0.434$ & $0.528^{+0.443}_{-0.442}$ & $2.772$ \\[0.5em]
	 		& optimized	& $-0.000\pm0.076$ & $0.002^{+0.060}_{-0.062}$ & $0.000\pm0.033$ & $0.002^{+0.026}_{-0.027}$ & $0.071$\\[0.3em]
	\hline\\[-0.9em]
	MIRROR & initial	& $0.082\pm0.405$ & $0.007_{-0.234}^{+0.473}$ & $0.133\pm0.251$ & $0.163_{-0.266}^{+0.189}$ & $0.463$ \\[0.5em]
	 		& optimized	& $-0.000\pm0.025$ & $0.001^{+0.021}_{-0.025}$ & $0.000\pm0.044$ & $0.005^{+0.029}_{-0.043}$ & $0.046$\\[0.3em]
	\hline
	\end{tabular}
\end{table*}

\begin{figure*}
	\centering
	\includegraphics[width=\textwidth, trim=0.2cm 0.2cm 0.2cm 2cm, clip=true]{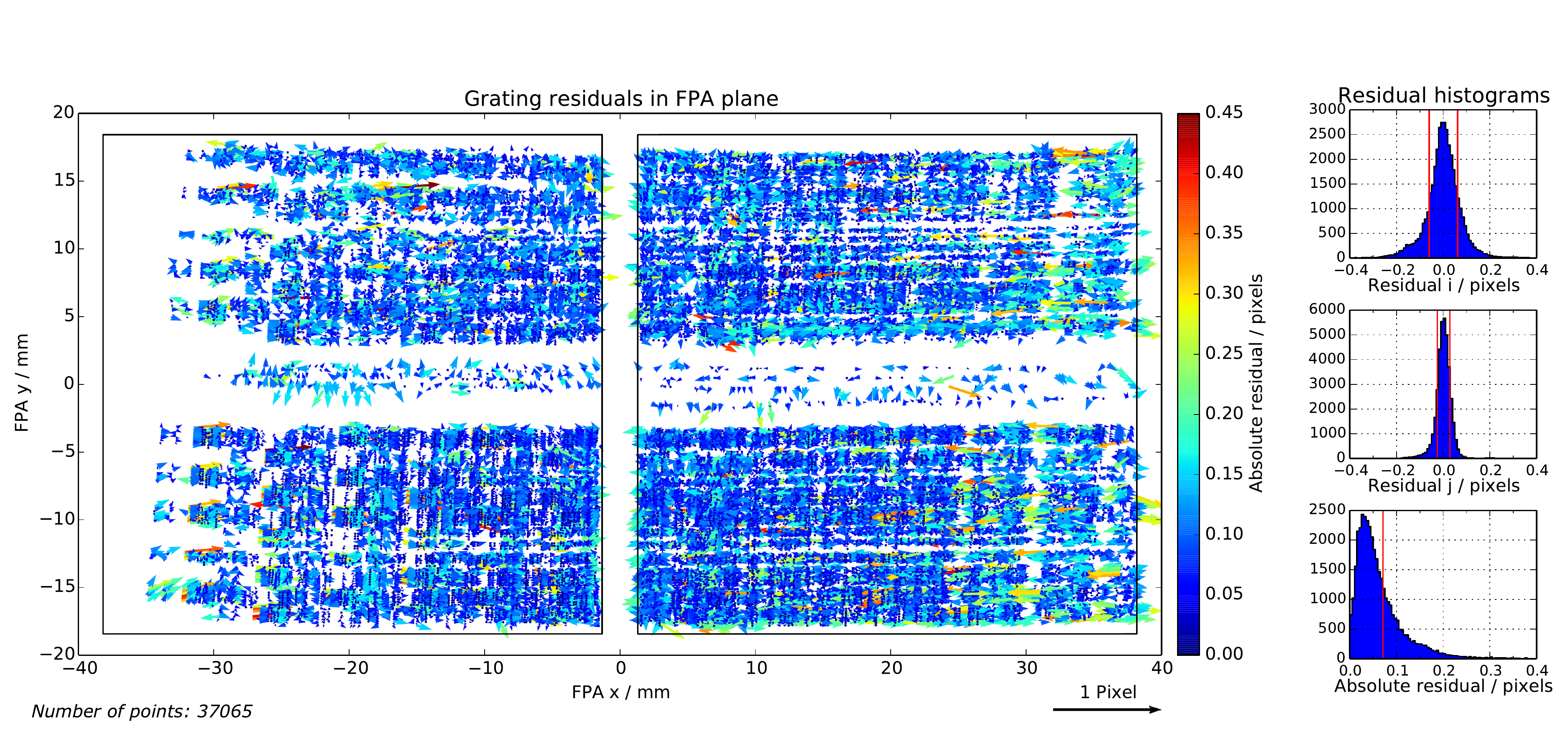}
	\caption{\label{fig:resgratings} Residuals of forward coordinate transforms from MSA to FPA for all of the gratings on the detectors.
	Left panel: Distribution in the detector focal plane.
	The spectra of the shutters are the wide bands on top and bottom; the stripes in the center are from the fixed slits.
	Right panel: Histograms of residuals in pixels in i and j direction (top, middle) and absolute residuals (bottom).
	Shown in red are the confidence limits of 68.27\% around the median at $-0.060$ and $+0.062$ (i), and at $-0.025$ and $+0.028$ (j), and the limit of 68.27\% of absolute residuals ($0.071$).}
\end{figure*}

\begin{figure*}
	\centering
	\includegraphics[width=\textwidth, trim=0.2cm 0.2cm 0.2cm 2cm, clip=true]{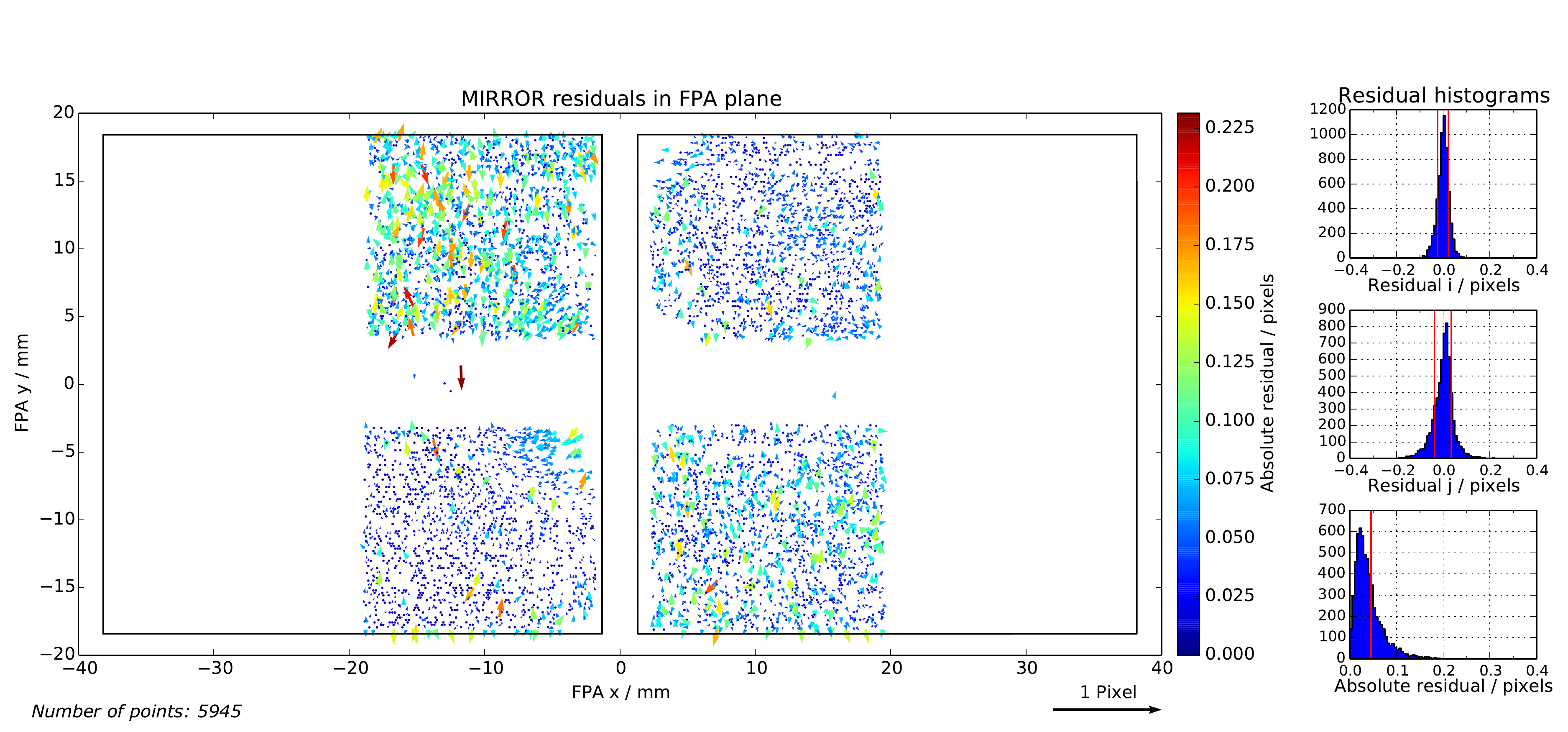}
	\caption{\label{fig:resmirror} Residuals of forward coordinate transforms from MSA to FPA for the imaging mode on the detectors.
	Left panel: Distribution in the detector focal plane.
	The four quadrants of the MSA are clearly visible, the five fixed slits are visible in the center area.
	Isolated shutters and slits are disturbed by bad pixels, in particular Q3 (top left).
	Right panel: Histograms of residuals in pixels in i and j direction (top, middle) and absolute residuals (bottom).
	Shown in red are the confidence limits of 68.27\% around the median at $-0.023$ and $+0.022$ (i), and at $-0.038$ and $+0.034$ (j), and the limit of 68.27\% of absolute residuals ($0.046$).}
\end{figure*}

In a second step, the positions of the fixed slits were optimized individually because their initial positions may not be optimal in combination with the new COL distortion, which was dominated by the MOS data in the previous optimization ($\approx 20\times$ more shutter references than for fixed slits).
Therefore, another fitting run was performed, only changing the SLIT positions in the MSA, using their reference data.
In a third step, we adjusted the alignment tilts of the two remaining gratings G140H and G235H to complete the forward transforms, taking the same minimization approach with their so far unused reference data.

The initial and final residuals on the FPA for fixed slits and MOS are listed in \autoref{tab:resopti}, the final ones are also shown in Fig.~\ref{fig:resgratings} for the gratings and Fig.~\ref{fig:resmirror} for the imaging mode.
As expected, the optimization brought the residuals on the detectors to an average of zero, while reducing the standard deviations of the gratings by a factor of about 13--21 from 0.43--1.6 to 0.033--0.076 px.
Some residuals are still on the order of 0.4 px, however, these correspond to single Argon lines, where the Gaussian fit was disturbed by nearby faint lines that were not detected in the line selection.
68.27\% of the values of the absolute residual (geometrical norm of both axes) are below 0.071 px (a limit similar to $1\sigma$ in a Gaussian distribution).
In imaging mode, the optimization was equally successful, reducing the residuals to an average of zero and their RMS by 6--16$\times$ to about 0.04 px.
Calculating the absolute residuals, 68.27\% of the values are lower than 0.046 px, which confirms the model accuracy.
Nevertheless, some of the shutter centroids are disturbed by bad pixels, especially in Q3 (lop left).
The large offset of the S200A1 (one of the left fixed slits) is likely caused by a low-QE pixel in the slit image and thus inaccurate centroid data.
The robustness of the optimization was confirmed by rerunning the procedure with different initial parameters, where we obtained the same final model.

The parametric model is not only used to project from the slits to detector, but also backward from FPA to the slit plane.
In fact, this direction is essential for the spectrum extraction and calculations for the target acquisition.
Therefore, two more steps were also necessary to optimize the backward transforms of the optical elements.
The CAM backward distortion was adjusted with the same reference data set as in the global fit, selecting the points G140M, G235M, G395M, G395H, and the MIRROR.
This optimization worked at the GWA exit plane in angular coordinates, since this is a pupil plane.
The MSA positions of the slits were transformed forward through the optimized COL and GWA.
The residuals were calculated as the difference between those projected slit position, and the positions of the detector reference points, transformed backward with the CAM model.
As in the previous steps, the backward distortion parameters of the CAM were optimized with a least-squares fit and $4\sigma$ clipping.
This approach was chosen to make the CAM backward transform optimal with respect to the COL forward transform, which is the same combination used during the wavelength calculation in the spectrum extraction (see Appendix \ref{AppMeeting}).

In the second step, the COL backward distortion was optimized with the identical reference data set.
To obtain a consistent transform from FPA to MSA, the reference points on the detector were propagated backward through the now optimized CAM and GWA elements to the GWA input plane.
The residuals were calculated between the detector references transformed further backward through the COL and the slit positions in the MSA plane.
The distortion parameters were again optimized with a least-squares fit and $4\sigma$ clipping.
With this step completed, the model was tuned to mimic the spectrographic part of the instrument for SLIT and MOS mode with all of the gratings and the MIRROR in both directions.

\begin{figure}
	\centering
	\includegraphics[width=\columnwidth, trim=0.6cm 0.1cm 0.6cm 0cm, clip=true]{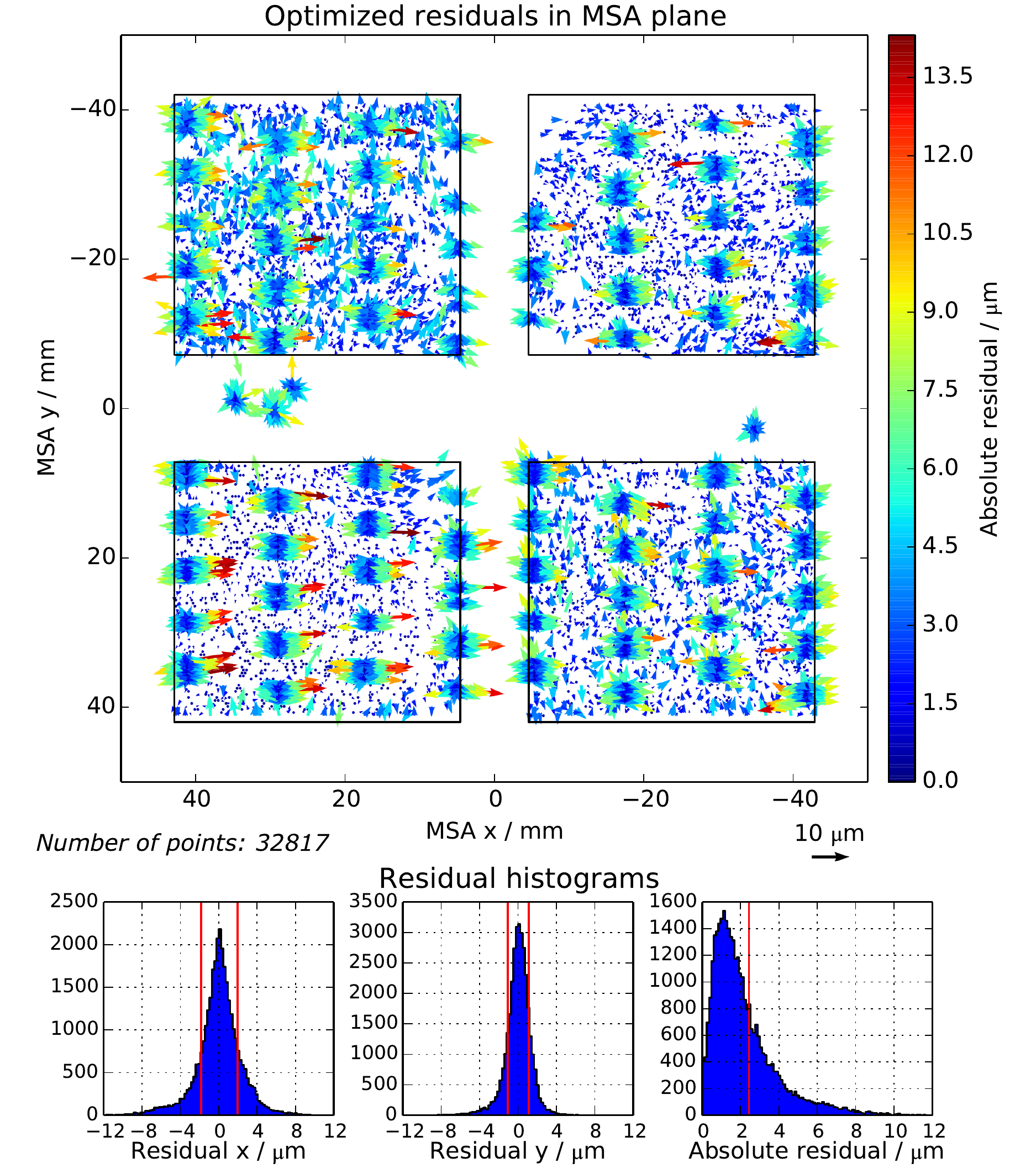}
	\caption{\label{fig:resmsa} Combined residuals of the backward transform from FPA to MSA for G140M, G235M, G395M, G395H, and the MIRROR. Top panel: Distribution in the MSA plane. The residuals of shutters used in the spectral modes (groups of the dashed slits) vary more strongly than those of the shutters from imaging mode (points spread in the quadrants). The physical direction of the axes are inverted to match the images of the detectors, since the reference frames are not following the optical projection. Bottom panel: Histogram of residuals in $x$ and $y$ direction (left, middle) and absolute residuals (right). Shown in red are the confidence limits of 68.27\% around the median at $-1.84$ and $1.98$ $\mu$m ($x$), and at $-1.07$ and $1.12$ $\mu$m ($y$), and the limit of 68.27\% of absolute residuals at $2.44$ $\mu$m.}
\end{figure}

The residuals of the complete backward transform from the detectors to the MSA plane are shown in Fig.~\ref{fig:resmsa} for the GWA elements used in the optimization (G140M, G235M, G395M, G395H, and the MIRROR).
The combined residuals are close to $0~\mu$m on average with a standard deviation and 68.27\% limit of absolute residuals around $2~\mu$m, which is equivalent to 1/38 of a shutter aperture width.
The imaging residuals alone are slightly more accurate with variations of around $1.4~\mu$m that are 1/54 of a shutter aperture width.
This is also apparent in the image, where the shutters of the spectral modes (groups of the dashed slits) show some larger residuals than those of the imaging mode (shutters spread in the quadrants).
However, the large residuals originate from problematic lines already seen in the forward direction (Fig.~\ref{fig:resgratings}) and are not representative of the average accuracy, as can be seen in the histograms.

The last subsystem missing in the spectrograph, the IFU, was handled independently, since its model offers the flexibility to adapt itself to the previously obtained parameters of the other components.
In the COL, the IFU virtual slits inject the light slightly outside the MSA field, but their position can be freely adjusted in the model, and in the CAM, the light mostly travels on paths common with the spectra of other modes.
At first, the virtual slits in the MSA plane were manually moved so that their projected outlines match their images on the detector and obtain an accuracy <1 px for the reference data creation.
Following this, 30 centroids of the virtual slit images and 918 spectral reference points were extracted in the same way as for the other slits, using exposures with the MIRROR and G395H grating and the Argon and continuum lamps.
As in the global forward optimization, the virtual slit positions were fitted by minimizing the residuals between the model predicted and measured positions on the detector.

Since the default reference data is averaged across the spatial direction, it does not contain any information about the slit rotation.
The IFU slice spectra are rather wide (30 pixels) and on this scale the optical distortion influences the tilt of the slit projection.
Therefore we performed another fitting round to adjust the rotation of the virtual slits.
Besides the references for the center, we extracted reference points from only the top and bottom half of the slice spectra, corresponding to traces at positions of $\pm0.25$ in the relative aperture and yielding additional information for the top and bottom half.
In this way, we obtained three reference points for each emission line at the relative positions$-0.25$, 0, and $+0,25$ inside the slice, which allowed us to verify  the slit rotation.
To determine the spectral coordinates, the Argon spectra were averaged only across the semislices, and the line positions $\lambda_m$ were measured as with a Gaussian fit as before.
However, in the spatial direction it is not possible to derive the centroids of only half a spectrum, so no polynomial function $P_j(i)$ could be constructed to calculate the pixel coordinates.
Nevertheless, after the preceding optimization, the model was already accurate in the spatial direction (the residuals were <0.1 px), therefore the spatial trace was calculated with the instrument model at the offset slice position.
Finally, in total we obtained 2778 points for the optimization, including the imaging centroids.

\begin{figure*}
	\centering
	\includegraphics[width=\textwidth, trim=0.2cm 0.2cm 0.2cm 2cm, clip=true]{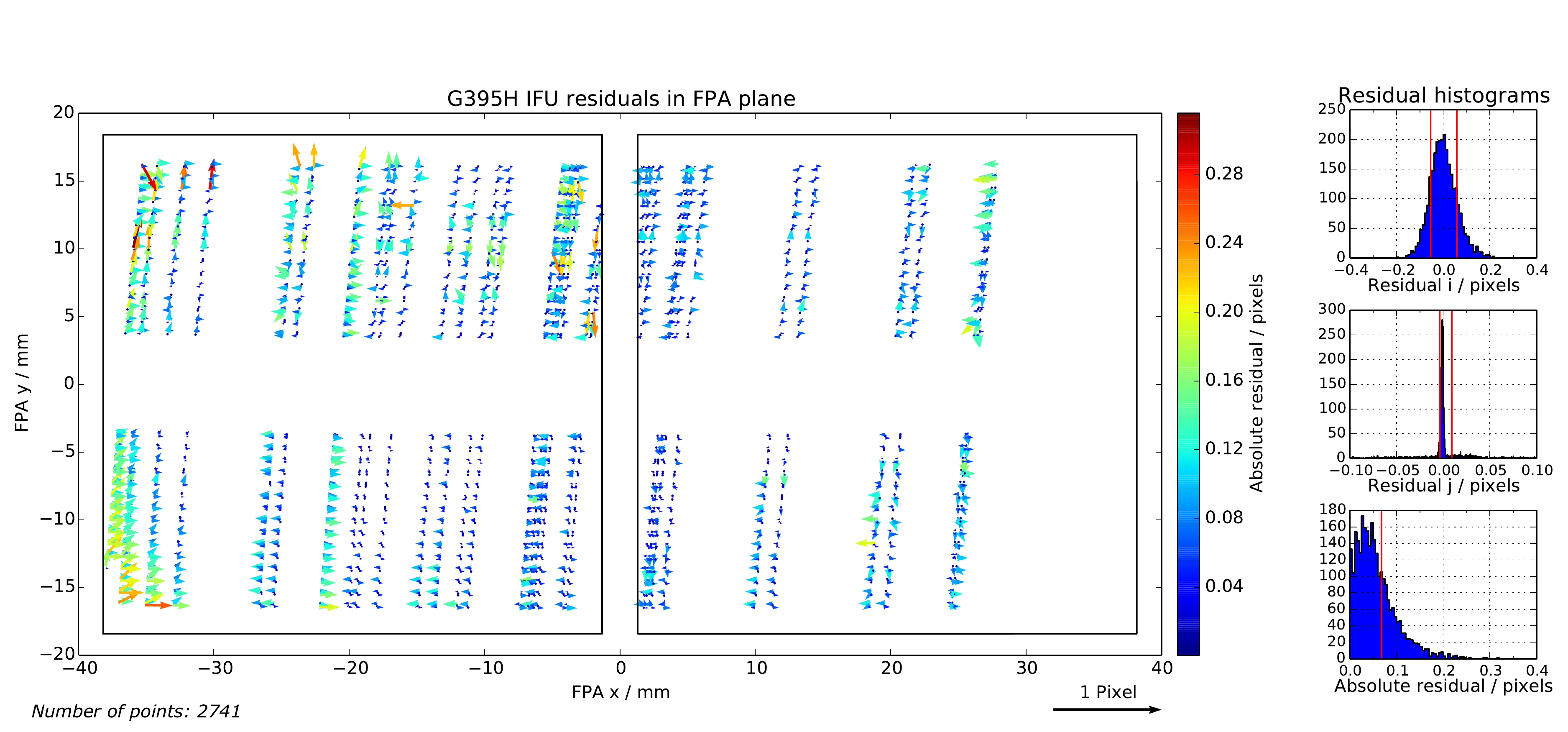}
	\caption{\label{fig:resifu} Residuals of transforms from IFU slices to FPA with G395H.
	Left panel: Distribution in the detector focal plane.
	The errors increase toward the lower left, where the CAM distortion was not influenced in the global fit and is not optimal for the IFU, but remain $<0.25$ px.
	Right panel: Histograms of residuals in pixels in i and j direction (top, middle) and absolute residuals (bottom).
	Shown in red are the confidence limits of 68.27\% around the median at $-0.054$ and $+0.058$ (i), and at $-0.004$ and $+0.009$ (j), and the limit of 68.27\% of absolute residuals ($0.068$).}
\end{figure*}

The final residuals of the forward transform with the IFU and the grating G395H are shown in Fig.~\ref{fig:resifu}.
The 68.27\% limit of absolute residuals is 0.068 px and similar to the SLIT and MOS modes; the errors increase only toward the left edge.
This is an area not covered by the MOS and SLIT data during the CAM distortion fit, therefore the optics model is not as accurate in that area.
Nevertheless the residuals there are sufficiently small with <0.2 px.
The spatial direction j is almost perfectly met, while the spectral direction and absolute residuals show a distribution similar to the other gratings in the global optimization step.
The IFU imaging residuals (not displayed) in the upper half of the left SCA 491 are disturbed by bad pixels and not reliable, similar to  the MOS mode.
The true accuracy of the centroids is likely better than the calculated 68.27\% absolute residual limit of 0.14 px.

%__________________________________________________________________
\section{Results}\label{Results}

To verify the self-consistency of the optimized model, we examined the resulting parameters for unrealistic values, which could mean that these values are degenerate with others or were not properly constrained during the fit sequence.
However, no unusual numbers were found in the results.

\begin{table}[b]
	\caption{Optimized geometrical parameters of the MSA model.}
	\label{tab:resMSA}
	\centering
	\begin{tabular}{cccc}
	\hline\hline
	Quadrant  	& Pitch x / $\mu$m & Pitch y / $\mu$m	&	Rotation / deg\\
	\hline
	1 & 104.97 & 203.90 & 0.03\\
	2 & 104.98 & 203.76 & 0.01\\
	3 & 104.99 & 203.92 & 0.01\\
	4 & 104.98 & 203.75 & 0.00\\
	\hline
	\end{tabular}
\end{table}

The MSA quadrants have pitches that differ by only $0.02\ \mu$m in the spectral and $0.15\ \mu$m in the spatial direction amongst them (see \autoref{tab:resMSA}).
The individual rotation angles are below 0.03\degr\ absolute.
The variations of fractions of microns and few arcminutes indicate the high consistency of the model.
The FPA was forced to end up in a symmetric model (as mentioned in Sect.~\ref{Assumptions}).
The fitted gap is 2.61 mm, and is very close to the nominal value of 2.75 mm, the rotation of SCA 492 is very low with 21.5 arcsec.

\begin{table}
	\caption{Optimized alignment tilt parameters of the gratings relative to the MIRROR surface.}
	\label{tab:resGWA}
	\centering
	\begin{tabular}{cccc}
	\hline\hline
	GWA & \multicolumn{3}{c}{Alignment tilt / arcsec}\\
	Element & $x$ & $y$ & $z$\\
	\hline
	G140H & 105.94 & 2.79 & 908.38\\
	G235H & 76.06 & 134.03 & 1248.71\\
	G395H & 74.53 & 109.93 & 860.25\\
	G140M & 112.35 & 40.49 & 846.99\\
	G235M & 116.38 & 134.86 & 810.66\\
	G395M & 205.06 & 47.26 & 877.13\\
	\hline
	\end{tabular}
\end{table}

The alignment angles of the gratings relative to the MIRROR surface are listed in \autoref{tab:resGWA}.
They generally show a large tilt of about 17\arcmin\ around the $z$-axis (perpendicular to the surface), which may be caused by freezing the CAM and FPA rotation, so that any other rotation in the field was transferred to the GWA parameters.
The tilt around the other axes $x$ (cross-dispersion movement) and $y$ (dispersion movement) are in the range of $\pm$100\arcsec\ and are consistent relative to each other.
Only G395M has a larger tilt of about 200\arcsec\ around $x$, which is evident in the exposures, where the spectra are located significantly lower on the detector compared to the other gratings.

The optical distortions of COL and CAM are largely dominated by low-order effects.
When re-deriving the paraxial magnifications and rotations from the new coordinate transforms, they would only change by <0.4\% and <0.015\degr\ compared to the as-built model.
Therefore, the optimized transforms are very close to the initial transforms from the Zemax prescription, giving us confidence in the fitted model as well as the as-built optical prescription.

\subsection{Spatial accuracy}\label{SpatCal}

\begin{table*}
	\caption{Residuals of calculated trace positions in the spatial direction compared to the measurements extracted with the optimized model. The data of each disperser were averaged over multiple spectral traces.}
	\label{tab:restraces}
	\centering
	\begin{tabular}{c c c c}
	\hline\hline
	GWA & \multicolumn{3}{c}{Residual / pixel} \\
	Element & SLIT & MOS & IFU\\
	\hline
	G140H & $0.054\pm0.060$ & $0.007\pm0.063$ & $0.049\pm0.091$ \\
	G235H & $0.050\pm0.091$ & $-0.018\pm0.088$ & $0.047\pm0.080$ \\
	G395H & $0.041\pm0.081$ & $0.004\pm0.065$ & $0.022\pm0.065$ \\
	G140M & $0.031\pm0.043$ & $0.007\pm0.037$ & $-0.004\pm0.077$ \\
	G235M & $0.022\pm0.038$ & $0.009\pm0.046$ & $-0.007\pm0.072$ \\
	G395M & $-0.012\pm0.068$ & $-0.007\pm0.049$ & $-0.011\pm0.097$ \\
	\hline
	\end{tabular}
\end{table*}

One specific way of testing the model is the verification of the spectrum trace positions.
To do this, using the optimized model, trace measurements were again extracted from the calibration data in the same way as described in Sect.\ \ref{Processing}.
The centroids in each pixel column of continuum spectra were fitted with a fourth-order polynomial with $4\sigma$ clipping.
The calculation was carried out for all of the gratings for SLIT and MOS. 
For the IFU, only the data of G395H had clean traces and were analyzed in the same way.
The exposures of other gratings that had not been used for the optimization were also processed, and clean areas of the spectra (no failed open shutter spectra) were selected manually.
Then, the slit centers were projected to the detector with the parametric model in the respective spectral range and the difference to the polynomials was computed for each column.
The average residuals in the spatial direction are listed in \autoref{tab:restraces}, along with the standard deviation.
The mean differences are typically <0.05 px and close to zero.
The scatter is <0.1 px and even smaller for the medium resolution gratings.
The IFU is similarly accurate in all of the gratings, although it has only been fitted with the G395H grating.

In imaging mode, the ability to reproduce the location of the slit images on the detector was already verified during the model optimization.
The large number of about 6000 randomly selected and manually inspected shutters used on input ensured a proper sampling of the FOV of NIRSpec.
This gives us confidence that the residuals shown in Fig.~\ref{fig:resmirror} and listed in \autoref{tab:resopti} are representative of all of the positions within the field of view.
Their standard deviations of 0.025--0.044 px represent an accuracy that is sufficient to meet the overall NIRSpec target acquisition error budget.

\subsection{Spectral accuracy}\label{WlCal}

The second important test of the parametric model is the extraction of spectra and verification of the spectral accuracy.
Therefore, the Argon lines from the input exposures were extracted to 1D spectra with the optimized model, the lines fitted with Gaussian functions, and their measured wavelengths compared to the true values.
The range below $1\ \mu$m was also examined for the band I gratings, which had not been covered by continuum spectra and thus was missing in the fit reference data.
In the IFU case, all of the slices for all of the dispersers were processed, as the impact of failed open shutters on the output was found to be negligible.
Lines that were known to be problematic during the model optimization were not taken into account.

\begin{table*}
	\caption{Residuals of extracted Argon emission lines comparing the measured line wavelength with the true wavelength. The data of each grating were averaged over multiple spectra and clipped with $4\sigma$.}
	\label{tab:reslines}
	\centering
	\begin{tabular}{*{7}{c}}
	\hline\hline
	 & \multicolumn{3}{c}{Residual / nm} & \multicolumn{3}{c}{Residual / pixel} \\
	Disperser & SLIT & MOS & IFU & SLIT & MOS & IFU \\
	\hline
	G140H & $-0.007\pm0.015$ & $-0.004\pm0.020$ & $0.017\pm0.024$ & $-0.031\pm0.063$ & $-0.019\pm0.083$ & $0.070\pm0.099$ \\
	G235H & $-0.024\pm0.028$ & $-0.019\pm0.036$ & $0.022\pm0.036$ & $-0.059\pm0.069$ & $-0.048\pm0.090$ & $0.055\pm0.090$ \\
	G395H & $-0.039\pm0.040$ & $-0.030\pm0.060$ & $0.000\pm0.033$ & $-0.058\pm0.059$ & $-0.044\pm0.090$ & $0.000\pm0.040$ \\
	G140M & $-0.006\pm0.025$ & $-0.013\pm0.033$ & $0.051\pm0.019$ & $-0.009\pm0.039$ & $-0.020\pm0.053$ & $0.080\pm0.030$ \\
	G235M & $-0.005\pm0.047$ & $-0.012\pm0.054$ & $0.076\pm0.049$ & $-0.005\pm0.044$ & $-0.011\pm0.050$ & $0.071\pm0.046$ \\
	G395M & $-0.006\pm0.080$ & $-0.034\pm0.085$ & $0.095\pm0.064$ & $-0.003\pm0.045$ & $-0.019\pm0.048$ & $0.053\pm0.035$ \\
	\hline
	\end{tabular}
\end{table*}

\begin{figure}
	\centering
	\includegraphics[width=\columnwidth, trim=0.6cm 0.2cm 0.6cm 0.9cm, clip=true]{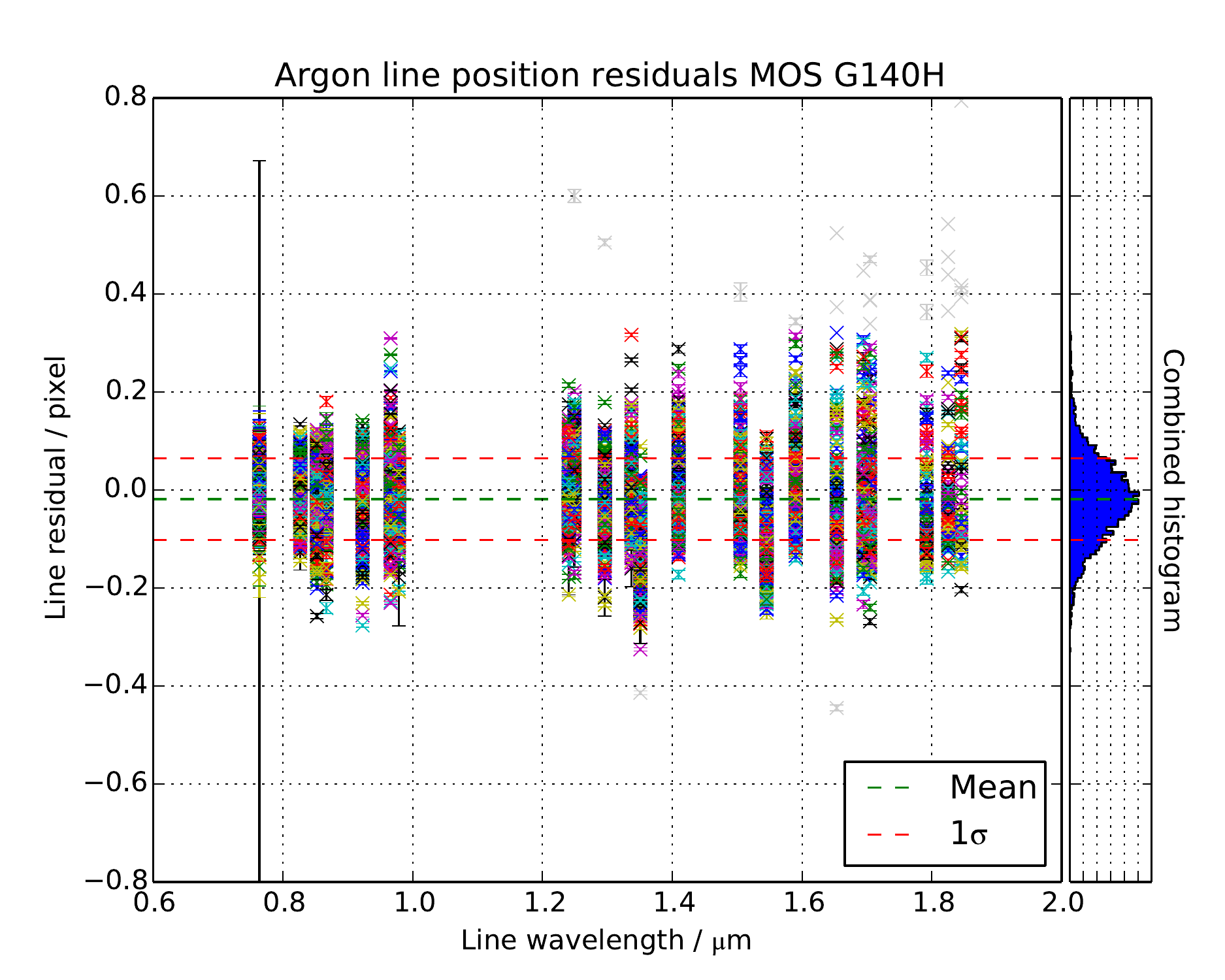}
	\caption{\label{fig:reslinesG140H} Extraction residuals of the Argon lines in the MOS data taken with filter F140X and grating G140H, given in detector pixels.
	In the main panel, each measured line position from each spectrum is plotted with the error from the fit.
	The mean residual is $-0.019$ px with a standard distribution of $0.083$ px.
	The points outside the 4$\sigma$-limit of the data are excluded as measurement errors.
	The right panel shows the distribution of the combined residuals of all lines.}
\end{figure}

The residuals of the modes and dispersers are listed in \autoref{tab:reslines}; they are individually clipped at $4\sigma$.
All of the gratings are accurately calibrated with a residual scatter around 0.08 px for the R2700, and 0.05 px for the R1000 dispersers, corresponding to 1/25 and 1/40 of a resolution element of 2 px, respectively.
As an example, the G140H MOS data are shown in Fig.~\ref{fig:reslinesG140H}.
The plot contains the residuals in pixels of each line from each spectrum with the error assigned during the Gaussian fit.
The majority of the points lie within $\pm$0.2 px and have negligibly small errors, and the overall standard deviation is 0.083 px.
The lines at wavelengths below 1 $\mu$m are also well calibrated, although the model has not been optimized with data in this spectral range.
Somewhat larger errors appear toward the longer end of the wavelength range, but can be attributed to random measurement errors.

%__________________________________________________________________
\section{Discussion and conclusion}\label{Discussion}

The results presented here show that the method we developed
yields an intrinsic spatial accuracy of the parametric model that is
better than 1/10 of a pixel. The model wavelength calibration is
equally accurate, showing RMS residuals equivalent to approximately
1/20 of a resolution element or, in other words, at the level of 5--7
km/s for all of the gratings and any of the NIRSpec modes (fixed slits, MOS,
and IFU). This is two times better than the formal standard deviation
allocation of 0.2 pixel (or 1/10 of a resolution element) specified
for this step in the overall wavelength calibration budget of
NIRSpec.

Reaching an even higher model precision would require more accurate
reference data.  The Gaussian fit of some emission lines is
problematic when their measured wavelength is influenced by blending
from nearby faint lines.  In Fig.~\ref{fig:reslinesG140H} this appears
as a bulk shift of the residual cloud of a few lines, for example, in
the line pair at $1.35~\mu$m.
Here the method reaches the limit of inspecting individual lines and
a fit of a line catalog could help to increase the model
accuracy further.

One limitation of this study is that the assessment of the wavelength
calibration accuracy is carried out, for the most part, on the same data
points from the same apertures that were used to
optimize the parametric model. During the limited time of the
calibration campaign, we did not acquire suitable additional data that
would allow us to assess, in a direct and clean manner, the wavelength
calibration accuracy from spectral lines from an independent set of
MOS apertures.

 We note, however, that the
optimization results in terms of the distortions of the COL and CAM
optics are extremely regular and smooth, as can be seen from
Fig.~\ref{fig:colcam_dist} in the
Appendix. This indicates that the optimization process is not driven
by the residuals of some individual lines/apertures, but a global
solution is obtained that reflects the macroscopic properties of the
optical system. Therefore we expect that comparable levels of accuracy
can be achieved for all of the MOS apertures on a different set of
unresolved spectral lines.  Indeed, as mentioned in
Sect.\,\ref{Results}, this is supported by the accuracy of the
wavelength calibration of the Argon lines in the range lower than
1~$\mu$m that (in the camera) sample different light paths from any of
the reference data.

So far we have not addressed the model accuracy for the large
square-shaped S1600A1 slit (intended for exoplanet transit
observations). This is because it was not practicable to fit Argon
lines to the large slit function, so the only reference data available
for this aperture are the slit image centroid coordinates.  However,
the centroid measurement is very robust thanks to the image extension
of 16$\times16~\mathrm{px}^2$, and in combination with the other elements of the
parametric model, and in particular the smoothness of the coordinate
transforms, it is expected to lead to an extraction of its spectral
data with an accuracy comparable to that of the other slits.

For a similar reason, we have not presented the results
of the wavelength calibration for the prism. The emission lines of the
Argon source cannot be used as wavelength reference in this case,
because they are heavily blended at the prism resolution of $\sim$100. Thus, a different data set is necessary for the calibration of
the prism. We plan to make use of data from three of the instrument
calibration sources of the CAA. These three sources employ Fabry-Perot
type interferometric filters that provide six well-defined transmission
features over Band I, II, and III. The accurate center of
gravity of these features at NIRSpec operating temperature
can be determined from the spectra obtained with
high resolution gratings, in a way bootstrapping the wavelength
calibration for the prism via that of the gratings presented here.
The results of this approach will be shown in a forthcoming paper.

The accuracy of the model transformations for the spectrographic part of
the instrument is a crucial part in the overall process of the NIRSpec
wavelength calibration, but it is not the only part. There are two
other important contributors to the overall wavelength calibration
budget: the calibration of the spectrum shift caused by the finite
repositioning accuracy of the grating wheel (see Sect. \ref{SpatCal}) and the
mixing of spatial and spectral information when our apertures are
significantly larger than the size of the point-spread function
(so-called slit effect).  Concerning the grating wheel-induced shift,
preliminary results from \cite{De-Marchi:2011_00} have shown that there is
a tight linear relation between the telemetry readings of the GWA
sensors and the tilt of a disperser element.  These calibration
relations are implemented during the spectral extraction to correct
for the actual disperser orientation. Current analysis using ground test
data shows that this correction leads to an additional contribution to
the uncertainty of the wavelength calibration of approximately 1/16
of a pixel, which is a fraction of the model uncertainties.

The calibration data presented here were acquired in 2013. In
the meantime, the MSA and FPA subsystems of the instruments have been
replaced. The quadrants of the new MSA have a smaller number of stuck
closed shutters (less than 10\%) and the new FPA unit features two new
H2RG detectors, which are not affected by the performance degradation
problem of the older model \citep{Rauscher+2014}. Therefore, the
instrument model needs to be updated to reflect the changes in the
optical geometry of these elements, which, however, are expected to be
small. For this purpose, we acquired a new set of imaging and
spectral reference data to be used for a new iteration of the model
optimization during the most recent cycle of cryogenic tests that the
instrument underwent in the Space Simulator Facility of NASA GSFC at
the end of 2015 and beginning of 2016. These tests involved all
  four scientific instruments onboard JWST in their flight
  configuration and integrated in the JWST Integrated Scientific
  Instrument Module (ISIM).
 
\begin{table}
	\caption{Planned updates of the NIRSpec parametric model
          following major milestones of the JWST project.}
	\label{tab:updates}
	\centering
	\begin{tabular}{lcc}
	\hline\hline
	 Dates & JWST Milestones & Model update \\
	\hline
        2015/16 & ISIM cryo campaign & Yes \\
        2016 & OTE-ISIM integration  &  - \\
        2016   & Vibration/acoustics tests & - \\
        2016/17 & OTE-ISIM cryo campaign & Yes \\
        2017/18 & Spacecraft integration & - \\
        2018 & Launch & - \\
        2018/19 & Commissioning & Yes\\
        $\ge$ 2019 & Normal operations & -\\
	\hline
	\end{tabular}
\end{table}

As the assembly of the space telescope proceeds, various
  environmental tests are planned for the different levels of the
  system integration. As summarized in \autoref{tab:updates},
  tied in with this overall schedule is our plan for three updates of
  the NIRSpec instrument model and further assessment of the
  wavelength calibration accuracy.  The first model update will take
  place once the set of recently acquired data have been reduced. In
  this case, an analysis of the spectral reference points from
  different MSA apertures and for a large set of GWA repositioning
  will be carried out to assess the comprehensive accuracy of the
  spectrograph wavelength calibration, over light paths not used for
  the model optimization itself and including the uncertainties of the
  GWA sensor calibration.

A new update of the model using the data acquired during
  the cryo-vacuum campaign at OTE-ISIM level will follow. This will allow us
  to assess the impact of the mechanical and thermal stresses that the
  instrument will undergo during the preceding series of
  vibration and acoustics tests (see \autoref{tab:updates}) on its
  overall alignment, and therefore, possibly, on the model
  parameters. The third and (likely) last update of the parametric
  model will take place soon after launch. We expect the mechanical
  and thermal environment of the telescope in space to be very stable
  and, therefore, once in operation, new model updates should not be
  necessary, although regular monitoring of the instrument
  wavelength solution will be performed.

Nevertheless, the model calibration presented here remains the
  keystone of all future model recalibrations. This is because,
  unlike during the 2013 test campaign, in the NASA testing facilities
  and in space we do not have access to an Argon source that provides
  us with a set of  independently and well-calibrated spectral
  references. The upcoming model updates will be carried out using
  NIRSpec internal calibration sources housed in the CAA. The
  wavelength of the emission and absorption features of these sources,
  however, was not known (from the manufacturer measurements and
  subsystem level tests) with the tenth-of-a-nanometer level of
  accuracy that we require for the NIRSpec wavelength calibration. To
  address this, during the 2013 test campaign, together with the Argon
  exposures we also acquired exposures of the internal
  sources and we used the model
  calibration presented here to obtain the wavelength
  calibration of the spectral features of the internal sources to the
  required accuracy.

Finally, for the sky observations and the target
acquisition, the calibration of the optical paths through the FORE
elements of NIRSpec has to be combined with the calibration of the
telescope optics and will be also carried out during the
  commissioning period, once the instrument is in space.
  Using the data acquired from observations of an astrometric field in
  imaging mode, the parameters of the FORE transform for each filter
  element in combination with that for the OTE will be
  determined. This step will complete the spatial and spectral
  calibration of NIRSpec prior to the start of any scientific activity.

In summary, the work presented here discusses the first and crucial
aspect of the NIRSpec wavelength calibration strategy: the
optimization of the instrument parametric model. We have developed a
procedure that uses the calibration data acquired for a limited subset
of the NIRSpec modes, and in particular only 1.5\% of  the 
quarter of a million NIRSpec slits, to derive a highly realistic model of the
instruments optical geometry.  This model allows us to compute
accurate light paths within the instrument and hence predict the
spatial and wavelength positioning of the spectra for all of the instrument
apertures and modes with an accuracy exceeding the formal requirement
by a factor of two.

%__________________________________________________________________
\begin{acknowledgements}

This work was funded in part by the Marie Curie Initial Training Network ELIXIR of the European Commission under contract PITN-GA-2008-214227.
\end{acknowledgements}

\bibliography{biblio}

%__________________________________________________________________
\begin{appendix}

\section{Coordinate transforms formalism}
\label{AppTransform}

\subsection{Transform between optical planes}\label{AppTrafoPlanes}

The coordinate transforms in the NIRSpec model use a paraxial
transform between the principal planes. Departures from the ideal
paraxial system are treated as distortions and modeled using a 2D
polynomial. 

The paraxial part of a forward coordinate transform is defined by the
magnification factors along the output axes $\gamma_x$ and $\gamma_y$,
the rotation angle of the coordinate system $\vartheta$, and the position
of the rotation center in the input and output reference frame
(x$_{0in}$, y$_{0in}$) and (x$_{0out}$, y$_{0out}$).  From the input
coordinates $(x_\mathit{in}, y_\mathit{in})$, the paraxial output
coordinates $(x_p, y_p)$ are given by
\begin{eqnarray*}
x_p & = & \gamma_x \cdot \left[(x_\mathit{in} - x_{0\,\mathit{in}})\cos(\vartheta) + (y_\mathit{in} - y_{0\,\mathit{in}})\sin(\vartheta)\right] + x_{0\,\mathit{out}}\ ,\\
y_p & = & \gamma_y \cdot \left[-(x_\mathit{in} - x_{0\,\mathit{in}})\sin(\vartheta) + (y_\mathit{in} - y_{0\,\mathit{in}})\cos(\vartheta)\right] + y_{0\,\mathit{out}}\ .
\label{eq:coot1}
\end{eqnarray*}

The optical distortion is applied in the form of a 2D polynomial of
order $n$, so the final output coordinates of the transform
$(x_\mathit{out}, y_\mathit{out})$ are
\begin{eqnarray*}
x_\mathit{out} & = & \sum_{i=0}^n \sum_{j=0}^{n-i} a_{i,j}(\lambda)\,x_p^i\,y_p^j\ ,\\
y_\mathit{out} & = & \sum_{i=0}^n \sum_{j=0}^{n-i} b_{i,j}(\lambda)\,x_p^i\,y_p^j\ ,
\label{eq:coot2}
\end{eqnarray*}
where the wavelength dependence is modeled by coefficients linear in wavelength $\lambda$ as follows:
\begin{eqnarray*}
a_{i,j}(\lambda) & = & \zeta_{x\,i,j} \lambda + \eta_{x\,i,j}\ , \\
b_{i,j}(\lambda) & = & \zeta_{y\,i,j} \lambda + \eta_{y\,i,j}\ .
\end{eqnarray*}

A backward transform is done in the reverse order.
At first, the distortion is removed and the paraxial coordinates are calculated
\begin{eqnarray*}
x_p & = & \sum_{i=0}^n \sum_{j=0}^{n-i} c_{i,j}(\lambda)\,x_\mathit{out}^i\,y_\mathit{out}^j\ ,\\
y_p & = & \sum_{i=0}^n \sum_{j=0}^{n-i} d_{i,j}(\lambda)\,x_\mathit{out}^i\,y_\mathit{out}^j\ ,
\end{eqnarray*}
where
\begin{eqnarray*}
c_{i,j}(\lambda) & = & \rho_{x\,i,j} \lambda + \sigma_{x\,i,j}\ , \\
d_{i,j}(\lambda) & = & \rho_{y\,i,j} \lambda + \sigma_{y\,i,j}\
\end{eqnarray*}
are the backward transform coefficients.
Then the input coordinates are
\begin{eqnarray*}
x_\mathit{in} & = & \frac{1}{\gamma_x} (x_p - x_{0\,\mathit{out}})\cos(\vartheta) - \frac{1}{\gamma_y} (y_p - y_{0\,\mathit{out}})\sin(\vartheta) + x_{0\,\mathit{in}}\ ,\\
y_\mathit{in} & = & \frac{1}{\gamma_x} (x_p - x_{0\,\mathit{out}})\sin(\vartheta) + \frac{1}{\gamma_y} (y_p - y_{0\,\mathit{out}})\cos(\vartheta) + y_{0\,\mathit{in}}\ .
\end{eqnarray*}

The paraxial description of the optical modules is based on the
nominal entrance and exit focal lengths, thereby in a transform between two
focal planes, for example, as in the FORE transform, where the coordinates are
the positions in the local reference frame in unit of length, the
magnification factors $\gamma_x$ and $\gamma_y$ correspond to the
ratio of the focal lengths $f$,
\begin{equation*}
\gamma_x = \frac{f_{out x}}{f_{in x}}, ~~~~~~~\gamma_y = \frac{f_{out y}}{f_{in y}}.
\end{equation*}
Transforms between a focal plane and a pupil plane (such as the COL
transform between the MSA and the GWA) are performed with similar 
formulas. In this case,
\begin{equation*}
\gamma_x = \frac{1}{f_{in x}}, ~~~~~~~\gamma_y = \frac{1}{f_{in y}},
\end{equation*}
and the output is a vector with a unitary $z$-component ($v_x, v_y, 1$).  The
light path through the dispersive elements of the GWA or the mirror,
however, is more easily computed using direction cosines.
The expression to convert from the transform result to
direction cosines $(\alpha, \beta, \gamma)$ is
\begin{equation*}
\begin{pmatrix}
\alpha\\
\beta\\
\gamma
\end{pmatrix}
=
\begin{pmatrix}
v_x/|v| \\
v_y/|v| \\
1/|v|
\end{pmatrix}
\label{eq:costodir}
\end{equation*}
with $|v| = \sqrt{v_x^2 + v_y^2+1}$.  The inverse operation to go from
direction cosines to the transform coordinates is, therefore,
\begin{equation*}
\begin{pmatrix}
v_x \\
v_y \\
\end{pmatrix}
=
\begin{pmatrix}
\alpha/\gamma\\
\beta/\gamma
\end{pmatrix}.
\label{eq:dirtocos}
\end{equation*}
 
\subsection{Transforms through the GWA elements}\label{AppGWATransforms}

When transforming the coordinates of a point in the MSA plane
to the FPA plane, one has to take into
account the effect of the mirror or the dispersive elements in the GWA
on the light path.
The specific coordinate transform for a light ray traversing a
given element in the GWA (whether the mirror, a grating, or the prism)
is applied after having rotated the ray coordinates at the collimator
exit to a reference frame with $z$-axis perpendicular the surface of the
given GWA element.
If the front surface is tilted  by $\Theta_y$, and the alignment tilt angles are $\vartheta_x, \vartheta_y, \vartheta_z$, then the first set of rotations applied to the incoming ray $\vec{i}_{in}$ in direction cosines $(\alpha_{in}, \beta_{in}, \gamma_{in})$ is
\begin{equation}
\begin{pmatrix}
\alpha^{in}\\
\beta^{in}\\
\gamma^{in}\\
\end{pmatrix}
= rot_{y}(rot_{z}(rot_{y}(rot_{x}(\vec{i}_{in}, \vartheta_x),\vartheta_y), \vartheta_z), \Theta_y)
\label{eq:rots1}\ ,
\end{equation}
where the suffix of the rotation functions denotes the axis around which the rotation is performed.
After application of the effect of the GWA element (see below), the outgoing vector $\vec{i}^{out}$ needs to be rotated back to the GWA output plane (that is CAM entrance plane), using the equation
\begin{equation}
\begin{pmatrix}
\alpha_{out}\\
\beta_{out}\\
\gamma_{out}\\
\end{pmatrix}
= rot_{x}(rot_{y}(rot_{z}(rot_y(\vec{i}^{out}, -\Theta_y), -\vartheta_z),-\vartheta_y), -\vartheta_x)\ .
\label{eq:rots2}
\end{equation}

The different types of GWA elements are modeled in different ways. In the case of the MIRROR, the light is only reflected, so we have the following wavelength-independent relation
\begin{equation}
\vec{i}^{out}_m
\begin{pmatrix}
\alpha^{out}_m \\
\beta^{out}_m\\
\gamma^{out}_m\\
\end{pmatrix}
= 
\begin{pmatrix}
- \alpha^{in} \\
- \beta^{in}\\
\gamma^{in}\\
\end{pmatrix}.
\label{eq:refl}
\end{equation}
For a grating, the chromatic coordinate transform is
\begin{equation}
\vec{i}^{out}_g
\begin{pmatrix}
\alpha^{out}_g \\
\beta^{out}_g\\
\gamma^{out}_g\\
\end{pmatrix}
= 
\begin{pmatrix}
- \alpha^{in}-k\lambda d \\
- \beta^{in}\\
\sqrt{1-(\alpha^{in}+k\lambda d)^2-(\beta^{in})^2}\\
\end{pmatrix},
\label{eq:grating}
\end{equation}
where $k$ is the grating order (nominal
value in NIRSpec case $k=-1$) and $d$ is the groove density.

Finally, for a light ray traversing the double-pass prism, the corresponding
coordinate transforms are given by the following equations.
The prism has a temperature and pressure dependent refraction index $n(\lambda, T, P)$, so 
Snell's refraction law through the front surface is
\begin{equation}
\vec{i}'
\begin{pmatrix}
\alpha' \\
\beta'\\
\gamma'\\
\end{pmatrix}
= 
\begin{pmatrix}
\alpha^{in}/n(\lambda,T,P) \\
\beta^{in}/n(\lambda,T,P)\\
\sqrt{1-(\alpha^{in}/n)^2-(\beta^{in}/n)^2}\\
\end{pmatrix}\ ,
\label{eq:p1}
\end{equation}
followed by the rotation to the reference frame of the prism back surface with the internal prism angle $\Theta_A$,
\begin{equation}
\vec{i}''
\begin{pmatrix}
\alpha'' \\
\beta''\\
\gamma''\\
\end{pmatrix}
= rot_y(\vec{i}', \Theta_A),
\label{eq:p2}
\end{equation}
the reflection on the back surface
\begin{equation}
\vec{i}'''
\begin{pmatrix}
\alpha''' \\
\beta'''\\
\gamma'''\\
\end{pmatrix}
=
\begin{pmatrix}
-\alpha'' \\
-\beta''\\
\gamma''\\
\end{pmatrix}\ ,
\label{eq:p3}
\end{equation}
going back to the surface frame
\begin{equation}
\vec{i}''''
\begin{pmatrix}
\alpha'''' \\
\beta''''\\
\gamma''''\\
\end{pmatrix}
= rot_y(\vec{i}''', -\Theta_A)\ ,
\label{eq:p4}
\end{equation}
and finally Snell's law through the front surface,
\begin{equation}
\vec{i}^{out}_p
\begin{pmatrix}
\alpha^{out}_p \\
\beta^{out}_p\\
\gamma^{out}_p\\
\end{pmatrix}
= 
\begin{pmatrix}
\alpha''''\times n(\lambda,T,P) \\
\beta''''\times n(\lambda,T,P)\\
\sqrt{1-(\alpha''''\times n)^2-(\beta''''\times n)^2}\\
\end{pmatrix}.
\label{eq:p5}
\end{equation}

\subsection{Meeting at the grating}
\label{AppMeeting}

\begin{figure*}
	\centering
	\includegraphics[width=\textwidth, clip=true, trim=0.2cm 0.5cm 0.2cm 0.5cm]{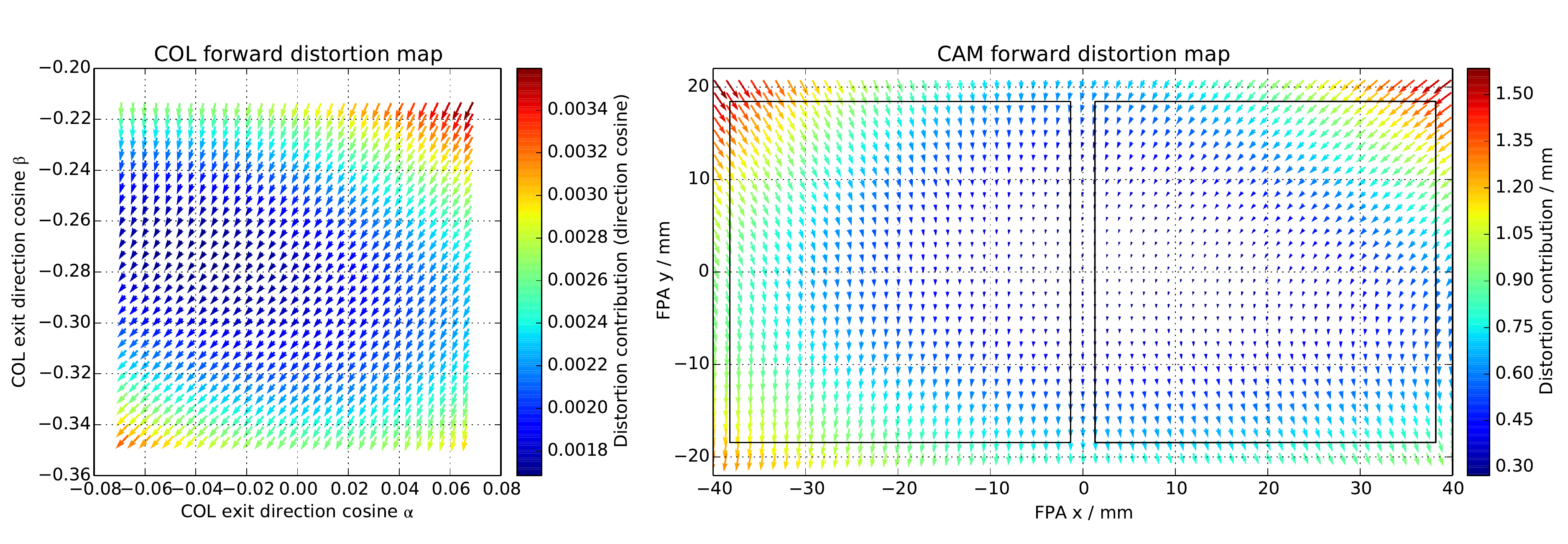}
	\caption{\label{fig:colcam_dist} Map of the optical distortions in the NIRSpec spectrographic part as obtained from the model optimization process. Left: collimator module, plotted in the exit plane. Right: camera module, plotted in the FPA plane. The outline of the two detectors is shown in black.}
\end{figure*}

One essential calculation in the extraction process is the derivation
of the 2D spectral coordinates (wavelength $\lambda$, spatial coordinate in the slit $d_y$) of a certain
position on the detector for a certain slit. Unlike the straightforward
transform from a slit to detector pixels at a given wavelength, this
operation cannot be achieved by simple geometrical transforms in
combination with the applicable dispersion law.  In an exposure
image, where the known parameters are the slit ID and the pixel FPA
coordinates, without knowing the wavelength, it is not possible to
calculate the transform at the disperser, and a brute force solution
for $\lambda$ and $d_y$ would
need a costly iterative optimization for each single point. However,
since the coordinate transforms of the COL and CAM optics are
achromatic, it is possible to solve for $\lambda$ and $d_y$ with an
 inverted linear interpolation, and a combination of the
individual transforms in the spectrograph.

First, the slit aperture is sampled at its center along the spatial
direction with a vector of points extending 5\% of the length beyond the
edges. This defines a set of coordinates at the MSA. These points are
then transformed through the collimator to the GWA plane, and rotated
to the disperser surface (just as in the usual transform), giving
pairs of two angular coordinates ($\alpha^{in}$, $\beta^{in}$), where
$\beta^{in}$ is primarily depending on $d_y$. On the scale of a
single slit, this dependency is approximately linear and an inverted linear
interpolation function $d_y(\beta^{in})$ can be computed.  Coming
from the other end of the instrument, the pixel coordinates at the FPA
are
transformed backward through the camera and rotated to the disperser
surface, giving two angular coordinates ($\alpha^{out}$,
$\beta^{out}$).  Since from the dispersion laws $\beta^{out} = -
\beta^{in}$, the coordinate $d_y$ can be readily
determined and, in turn, used to calculate $\alpha^{in}$ with another forward transform from the slit.
Finally, the coordinate $\lambda$ can be computed
by inverting the dispersion law. For the grating, this is simply
given as
\begin{equation}
\lambda = \frac{-\alpha^{in}-\alpha^{out}}{k d}\ ,
\end{equation}
where $n$ is the grating order and $d$ is the groove density.
For the prism, after computing the value of the refraction index for a
number of wavelengths within the nominal wavelength range, an inverted
interpolation relation $\lambda(n)$ can be set up and used to derive
the value of $\lambda$ from the value of $n$, which is given by
\begin{equation}
n = \sqrt{\frac{\alpha^{out}+\alpha^{in}(1-2\sin^2
      \Theta_A)}{2\sin\Theta_A \cos\theta_A}+\alpha^{in}{}^2+\beta^{in}{}^2}\ ,
\end{equation}
where $\Theta_A$ is the internal prism angle.

\section{Optical distortions of COL and CAM}

Fig.~\ref{fig:colcam_dist} shows the maps of the distortions of COL and CAM as obtained from the model
optimization process. Their smoothness and regularity indicates that the optimization process is not driven by individual
lines or MOS apertures.

\end{appendix}

\end{document}